\documentclass[layout=onecolumn, fontsize=10]{article}

\usepackage[utf8]{inputenc}
\usepackage[T1]{fontenc}
\usepackage[english]{babel}

\usepackage{amsfonts}
\usepackage{amsmath}
\usepackage{cleveref}

\usepackage{graphicx}
\usepackage{wrapfig}
\usepackage{pdfpages}
\usepackage{tabularx}
\usepackage{multirow}
\usepackage{tabulary}
\usepackage{booktabs}
\usepackage{subcaption}
\usepackage{xr}
\usepackage{multicol}
\usepackage{authblk}
\usepackage[backend=biber,style=chem-acs]{biblatex}
\usepackage[papersize={8.5in,11in},left=2cm, right=2cm, top=3cm, bottom=3cm]{geometry}

\newcommand{\angstrom}{\mbox{\normalfont\AA}}
\newcommand{\etal}{{et al.\,}}
\let\oldmaketitle\maketitle
\let\maketitle\relax

\bibliography{literature}

\author[1,a]{Mirko Fischer}
\affil[1]{Institiute of Physical Chemistry, Westfälische Wilhelms-Universität Münster, Corrensstraße 28/30, Münster 48149, Germany}

\author[1]{Andreas Heuer}

\author[2]{Diddo Diddens}
\affil[2]{Forschungszentrum Jülich GmbH, Helmholtz Institute Münster, Corrensstraße 46, Münster 48149, Germany}

\affil[a]{Corresponding author: mirko.fischer@uni-muenster.de}

\title{\textbf{Structure and transport properties of poly(ethylene oxide) based cross-linked polymer electrolytes - A Molecular Dynamics Simulations study}}
\date{}

\begin{document}
	
\twocolumn[
\begin{@twocolumnfalse}
\oldmaketitle
\begin{abstract}
We present an extensive molecular dynamics (MD) simulation study of poly(ethylene oxide) (PEO) based densely cross-linked polymers, focussing on structural properties as well as the systems' dynamics in the presence of lithium salt. Motivated by experimental findings for networks with short PEO strands we employ a combination of LiTFSI (Lithium bis(trifluoromethanesulfonyl)imide) and LiDFOB (Lithium difluoro(oxalato)borate). Recently, it has been shown that such multi-salt systems outperform classical single salt systems (Shaji \etal, \textit{Energy Storage Materials}, \textbf{2022}, 44, 263). To analyse the microscopic scenario we employ an analytical model, originally developed for non-cross-linked polymer electrolytes or blends (Maitra \etal, \textit{Phys. Rev. Lett.}, \textbf{2007}, 98, 227802 and Diddens \etal, \textit{J. Electrochem. Soc.}, \textbf{2017}, 164, E3225-E3231).
Excluding very short PEO strands, the local dynamics is only slightly restricted compared to linear PEO and is not significantly dependent on the network structure. The transfer of lithium ions between PEO chains and the motion along the polymer backbone may be controlled through the employed salt.

\vspace{0.4cm}
\centering
\includegraphics{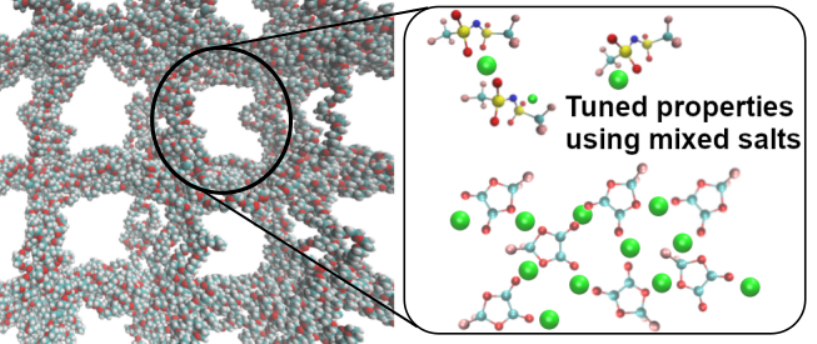}
\end{abstract}

\end{@twocolumnfalse}]
 
	\section{Introduction}
	
	Batteries play an important role in our society. In the future, this technology will be even more important and will be needed for various mobile applications. Therefore batteries must satisfy different criteria, such as a high capacity, a high mechanical stability and a high conductivity of the electrolytes.
	Besides the broadly used lithium-ion-batteries, lithium-metal-batteries are promising \cite{Liu.2020, Hwang.2019}. However, lithium-metal-batteries come with some problems that have to be solved. Metallic lithium is very reactive and while charging inhomogeneous deposition of metallic lithium, so-called dendrites, at the electrodes takes place \cite{Cheng.2017, Xu.2014, Arya.2017}. This leads to safety risks.
	Instead of classical liquid electrolytes, solid polymer electrolytes (SPEs), which suppress dendrites due to their mechanical stability, can be used \cite{Long.2016}. 
	Unfortunately, the conductivity of polymer electrolytes is still too low \cite{Nair.2019b}. In the case of the often used poly(ethylene oxide) (PEO) there is a strong coupling between PEO and lithium which slows down the transport of lithium ions ($\mathrm{Li^+}$). Nevertheless, a certain interaction between the polymer and the salt is needed to dissolve lithium salts in PEO. Moreover, PEO crystallizes at low temperatures that are desired for many batteries. This slows down the lithium ion transport even more \cite{Bresser.2019}. 
		
	\begin{figure*}[t]
		\centering
		\includegraphics[width=\textwidth]{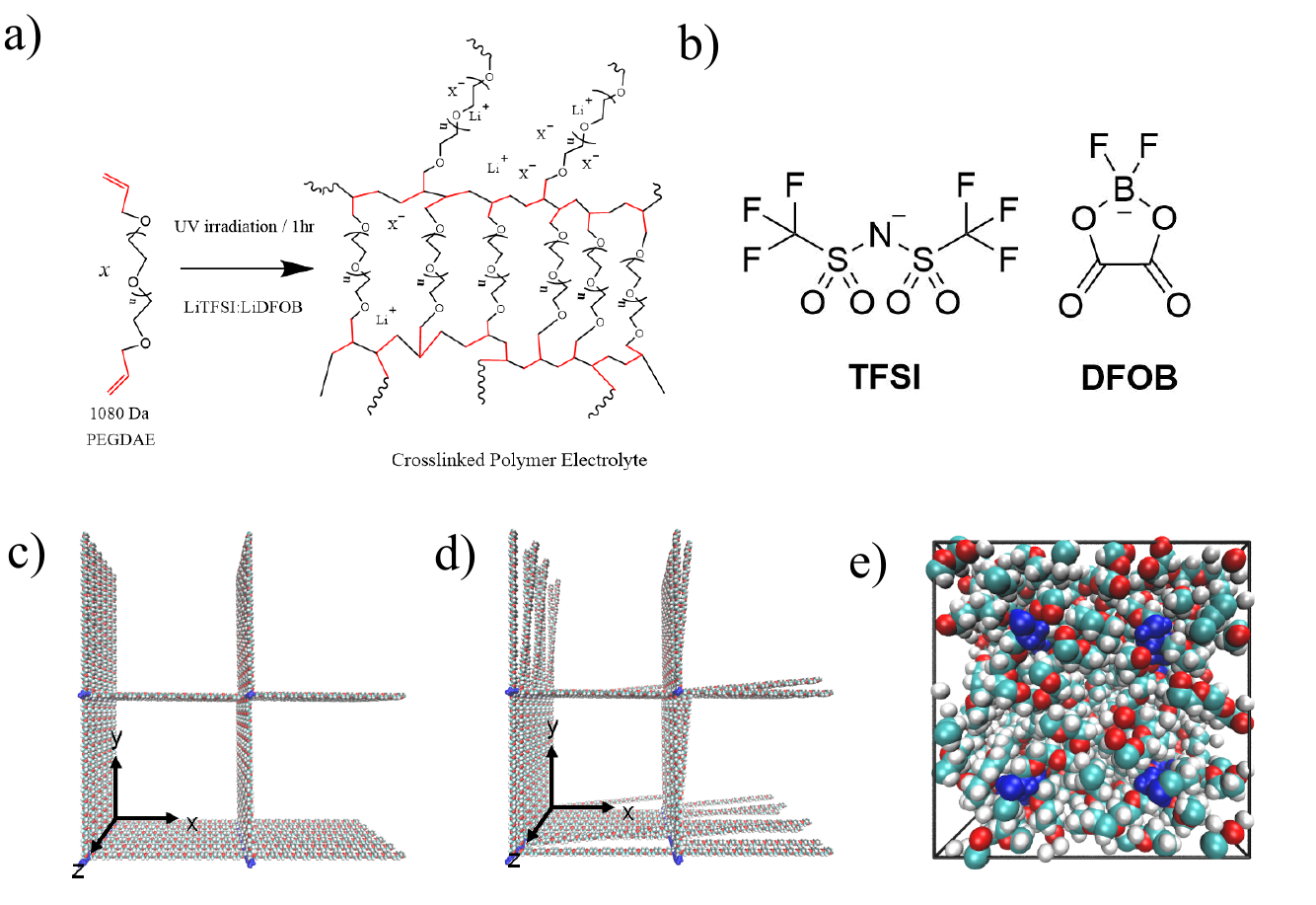}
		
		\caption{a) Reaction used experimentally for cross-linking PEO polymer chains to a network structure, b) lithium salt anions employed in this work, c) initial idealized network structure before equilibration. Each PEO chain is connected at both of its ends to a PE chain, d) initial network structure with network defects. Randomly 25\% of the PEO chains are connected only at one of their end to a PE chain, e) Equilibrated network structure. The PE chains are highlighted in blue.}
		\label{structures}
	\end{figure*}
	
	One solution is to use cross-linkers to form cross-linked polymers. Their stability is improved and crystallization is suppressed, because the cross-linkers keep the polymer disordered.
	In fact, it has be shown that a LiTFSI-LiDFOB dual-salt is a promising candidate for SPEs in lithium-metal-batteries \cite{Liu.2020}. In their recent work Shaji et al. \cite{Shaji.2022} investigated various dual salt systems with LiTFSI employed as the major salt and LiDFOB as the secondary salt in different ratios. The dual salt systems outperformed the single salt LiTFSI system in terms of ionic conductivity, electrochemical stability, lithium ion diffusion coefficient and lithium ion transference number. 
	Preliminary MD simulations could show that the choice of the salt affects both polymer and ion dynamics \cite{Shaji.2022}. However, the network structure was simplified as a PEO melt with fixed chain ends.
	In this work, we substantially extend these preliminary findings by modeling the complete network structure, and unravel the molecular transport mechanism in more detail. We do not only analyse the effect of cross-linked PEO polymers on the lithium ion transport, but also the effect of both lithium salts within the polymer electrolytes.
	As polymers we used PEO cross-linked with polyethylene (PE). The chemical reaction underlying the formation of the network is shown in Figure \ref{structures}a. The cross-linked polymer can be synthesized by UV-induced free radical photopolymerization of allyl-terminated PEO chains as (poly(ethylene glycol))diallyl ether (PEGDAE)\cite{Shaji.2022}.  
	While experimentally LiDFOB is frequently used as an alternative to LiTFSI, the former has only rarely been investigated by MD simulations \cite{Han.2013}.

	This paper is structured as following. First, we briefly explain the transport model used for the analysis of the data. Then we present the structural properties of the polymer such as polymer orientation and coordination with special regards to the effect of cluster formation. Next, we discuss dynamics, including relaxation times as defined in the transport model and the mean squared displacement of different species as well as the lithium ion diffusion coefficient. We conclude with a short comparison to non-ideally cross-linked polymers as well as non-cross-linked polymers.
	
	\section{Theoretical Background}
		
	\subsection{MD-Simulations}
	\label{sec_MD_simulations}
	
	All simulations have been performed with the GROMACS-2019.3 package \cite{Abraham.2015}. The molecular interactions were described by the OPLS-AA force field \cite{Jorgensen.1996} for PEO as well as the CL\&P force field \cite{CanongiaLopes.2004,CanongiaLopes.2004b,CanongiaLopes.2012} for TFSI.
	
	The partial charges of DFOB were derived via the ChelpG method \cite{Breneman.1990} using the Gaussian 16 package \cite{M.J.FrischG.W.TrucksH.B.SchlegelG.E.ScuseriaM.A.RobbJ.R.CheesemanG.Scalmani.2016} on DFOB geometries optimized at the HF/6-31G(d) level of theory according to the general procedure of the parame\-trization of the CL\&P force field \cite{CanongiaLopes.2004,CanongiaLopes.2004b,CanongiaLopes.2012}. The partial charges themselves were computed at the MP2/cc-pVTZ level of theory \cite{Frisch.1990,Dunning.1989}. We find (unscaled) charges of 1.06\textit{e} at the boron atom (\textit{e} being the elementary charge), -0.49\textit{e} for the fluorine atoms, 0.67\textit{e} for the carbon atoms, -0.59\textit{e} for the oxygen atoms connected to boron as well as -0.62\textit{e} for the carboxyl oxygen atoms. Bonded DFOB parameters were adopted from the OPLS-AA parameters of $\textrm{BF}_4^-$, acetate, or ethers, improper terms from the standard OPLS-AA aromatics. 
	The parameters for $\textrm{Li}^+$ were taken from \cite{Aqvist.1990}.
	
	To account for polarization effects, all partial charges have been scaled by a factor of 0.8 as reported elsewhere in the literature \cite{Leontyev.2010,Molinari.2018,Thum.2020}
			
	Different systems were created to analyse the effect of two lithium salts and the PEO chain lengths on the systems properties.
	
	Systems with each of the pure salts \mbox{LiTFSI} and \linebreak\mbox{LiDFOB} and a dual salt system with a ratio of \linebreak\mbox{$n(\mathrm{TFSI})/n(\mathrm{DFOB}) \approx 1.43$} were simulated, motivated by the systems with optimal performance in the experiments by Shaji et al \cite{Shaji.2022}.
	To analyse the effect of chain lengths, systems with 12 (64 chains), 24 (48 chains) and 36 (64 chains) monomers per chain were studied. In all systems the ratio of Li:EO is approximately 1:14. Corresponding to the system size, the systems were set up with 54, 82 and 163 ion pairs. 
	Unfortunately, the real structure resulting from the polymerization reaction is unknown, which requires to make some simplifications. First, each PEO chain is connected at its ends to a PE chain and second, the chains are regularly ordered and stretched in the initial structure, which leads to a lattice structure with large cavities (Figure \ref{structures}c). In the simulated systems, all four PE chains are orientated along the z direction. The PEO chains are orientated in the x and y direction (see Figure \ref{structures}c). Before equilibration, the lithium salts LiTFSI and LiDFOB (Figure \ref{structures}b) were placed in the cavities of the lattice structure.
	To keep the computational cost at a minimum, the systems were chosen as small as possible yet still representative.
	Due to the slight anisotropy of the system, which persists upon equilibration to constant density, our simulation boxes are only approximately cubic with dimensions of \mbox{3.5 x 3.5 x 3.5 $\mathrm{nm}^3$} to \mbox{5 x 5 x 5 $\mathrm{nm}^3$}. The chosen size enabled us to simulate 800\,ns to 1000\,ns at still acceptable computational costs.
	Note that larger systems with 160 PEO chains and 272 ion pairs (24 monomers per chain) were additionally simulated to make sure smaller systems are representative (for further reading see supporting information).
	Moreover, to understand which effects imperfectly formed networks may have on the systems dynamics, systems were simulated in which 25 \% of PEO-chains are only connected to one PE chain (see Figure \ref{structures}d). Because the structure in these systems changes slightly, we had to enlarge the size of the simulated systems. The structural changes are manifested in a reduced distance between the PE chains, which are also slightly more stretched than in systems without network defects. Systems with 64 PEO chains with a length of 24 monomers per chain and 110 ion pairs were used in this case.  
	Finally, a system without network structure was set up with 40 PEO chains and a length of 36 monomers per chain. 60 TFSI and 42 DFOB anions were placed in this system.
		
	Electrostatic interactions have been treated by the particle mesh Ewald summation \cite{Darden.1993,Essmann.1995} with a cutoff radius of 1.6 nm using a grid spacing of 0.1 nm and 6-th order interpolation. For Lennard-Jones interactions a cutoff radius of 1.6 nm was used.
	
	The systems were relaxed over a period of 1 ps with a time step of 0.005 fs with a reduced cutoff radius of 1.2 nm at a temperature of \mbox{$T=450 \mathrm{~K}$} and a pressure of \mbox{$p = 100 \mathrm{~bar}$}. 
	Subsequently, the systems were equilibrated in the NpT-ensemble over a period of 60 ns with a timestep of 2 fs. Afterwards, the pressure was changed to \mbox{$p = 1 \mathrm{~bar}$}. Finally, production runs over a period of 400 ns for the larger reference systems and 800 - 1000 ns for the other systems have been performed, respectively.
	Temperature and pressure were maintained by the Berendsen thermostat and barostat \cite{Berendsen.1984} during equilibration and by the Nos\'{e}-Hoover thermostat \cite{Nose.1984,Hoover.1985} and Parrinello-Rahman barostat \cite{Parrinello.1981,Nose.1983}  during the production runs. Time constants of 0.5\;ps and 2.0\;ps were used for the thermostats and barostats, respectively. The pressure coupling was performed semi-isotropically.  
	Periodic boundary conditions were applied in all three dimensions. H-bonds were constrained by the LINCS algorithm \cite{Hess.1997} during equilibration and production run. For integration, the leap frog algorithm was used.
	
	\subsection{Analytical model}
	\begin{figure}[t]
		\centering
		\includegraphics[width=\linewidth]{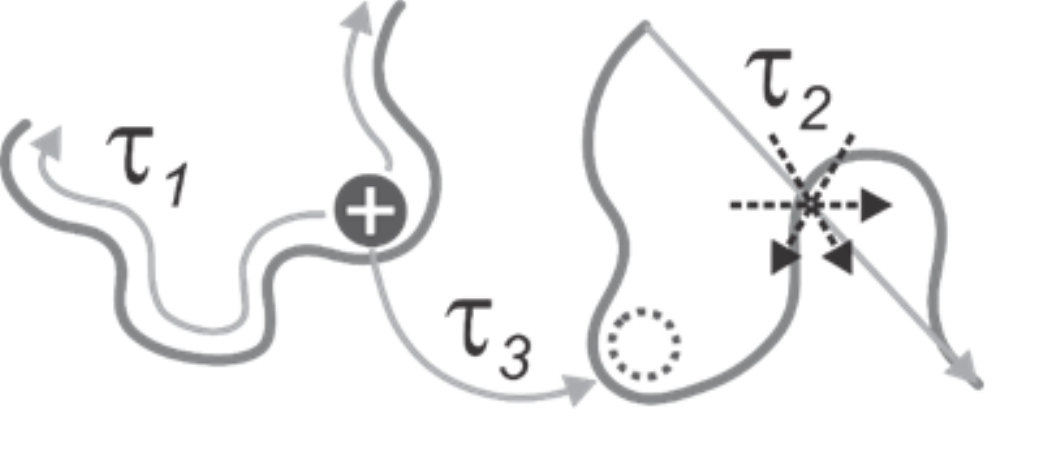}
		\caption{Illustration of the microscopic transport model for lithium ions. Each mechanism can be quantified by a corresponding time scale. Reprinted with permission from \cite{Diddens.2013}. Copyright 2013 American Chemical Society.}
		\label{mechanism}
	\end{figure}

	To analyse the results of our MD-simulations, we apply an analytical lithium ion transport model, developed for PEO based electrolytes \cite{Maitra.2007,Diddens.2010,Diddens.2017}. In the past, it was successfully applied to different systems containing for example PEO/LiTFSI systems and ternary polymer electrolyte-ionic liquid mixtures. However, it has not been applied to ion conducting polymer networks, which we address in this article.
	The model is based on the Rouse-Modell \cite{Rouse.1953,Rouse.1998} (which takes the polymer dynamics into account) and the Dynamic Bond Percolation-model \cite{Druger.1983} (DBP, which takes random jump processes within the polymer matrix into account). 
	In detail, three different mechanisms must be considered, which are shown in Figure \ref{mechanism}.  
	To quantify the contribution of the individual mechanisms to the total lithium ion dynamics, corresponding relaxation times $\tau_\mathrm{i}$ can be calculated.  
	Within the model, lithium ions are assumed to be mainly coordinated to the polymer chains, which typically is the case for PEO-based electrolytes. As a first mechanism, lithium ions are able to move along a chain. The relaxation time $\tau_1$ can be interpreted as the time a lithium ion needs to explore the entire chain. 
	Second, the dynamics of the lithium ions is affected by the dynamics of the PEO chains above the glass transition temperature so that the lithium ions move together with the polymers. For short chains this motion consists of the center-of-mass motion and the segmental motion, whereas the former becomes negligible for long chains. The segmental motion can be quantified by relaxation time $\tau_2$ and can be interpreted as an effective Rouse time if the bound segments. 
	Third, a lithium ion, coordinated to a specific polymer chain, can be transferred to another polymer chain. We also call this process jumping of a lithium ion. The mean time between two jumps is denoted as $\tau_3$. 
	Note that these jumping events are treated like renewal events in a Random Walk model. After a jump the motion of a lithium ion becomes uncorrelated from its past.
	Based on theoretical considerations, it was found for linear polymer chains that $\tau_1, \tau_2 \propto N^2$ and $\tau_3 \propto N^0$, where $N$ is the number of monomers of the polymer chains \cite{Chattoraj.2015}. Additionally, for short chain lengths, the center-of-mass motion of the polymer chains significantly contributes to the overall dynamics. However, in our system the ends of the polymer chains are constrained due to the network structure and therefore we expect that there is no center-of-mass motion of the polymer chains. Thus, we are in the limit of infinitely long chains even for very short chains. 
	In case of ionic liquids and all systems, in which a significant fraction of the lithium ions are not coordinated to any polymer chain, a fourth mechanism becomes relevant.
	Lithium ions can move cooperatively with anions, when lithium ions are not coordinated to any polymer chain. Furthermore, shuttle molecules such as oligo-ether functionalized ionic liquid cations can be used to decouple the lithium ions from the slow polymer chains \cite{Atik.2021,Diddens.2017}. In principle, also anions can be used for this decoupling.
	As for mechanism 3, the motion of a lithium ion becomes uncorrelated to its past, when it is transferred from a polymer chain to anions/shuttles or vice versa. 
	Consequently, a correction term to the polymer based diffusion coefficient $D_{\mathrm{Li}}^{\mathrm{PEO}}$, predicted by the model, has to be incorporated. Assuming that the dynamics of lithium ions coordinated to PEO and anions is statistically independent, the total diffusion coefficient is then given as $D_{\mathrm{Li}}^{\mathrm{tot}} = (1-p_{\mathrm{PEO}}) \cdot D_{\mathrm{Li}}^{\mathrm{anion}}+ p_{\mathrm{PEO}} \cdot D_{\mathrm{Li}}^{\mathrm{PEO}}$, implying that the lithium ions are either coordinated to the polymer chains or exclusively to the anions. $p_\mathrm{PEO}$ is the percentage of lithium ions that are coordinated at least to one PEO chain and $D_\mathrm{Li}^\mathrm{anion}$ is the diffusion coefficient of lithium ions that are coordinated to anions.
	To compare the results of our simulation to experiments, we can furthermore calculate an approximation for the transference number $t_{\mathrm{+}}=D_{\mathrm{Li}}^{\mathrm{tot}}/(D_{\mathrm{Anion}}+D_{\mathrm{Li}}^{\mathrm{tot}})$. Note that this expression neglects the presence of dynamical ion correlations and is also termed ion transport number \cite{Wohde.2016}.
	
	\section{Results}
	
	\subsection{Network structure}
	\label{Sec:network_structure}
	
	\begin{figure*}[t]
		\centering
		\includegraphics[width=\linewidth]{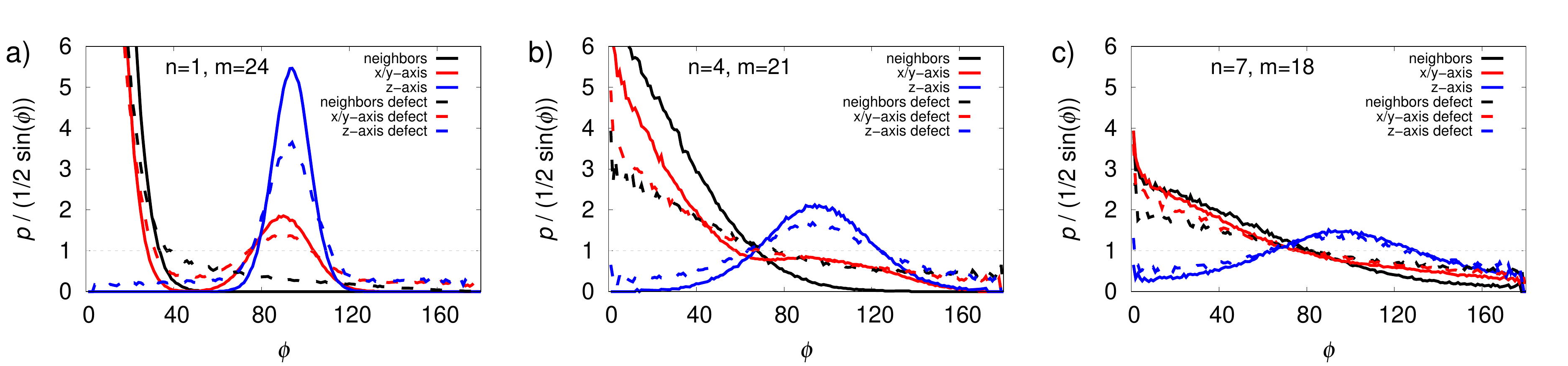}
		\caption{The probability distributions of the angles between the PEO chains and unit vectors of the individual spatial dimensions as well as the angles between a vector in a PEO chain and a vector in a neighbouring PEO chain are shown for a system with a PEO chain length of 24 monomers. The curves have been divided by $\frac{1}{2}\sin(\phi)$. The vectors between the monomers $n$ and $m$ are calculated a) from the first EO in chain to the last EO in chain, b) from 4th EO to 21st EO (90\% of lithium ions are in this region) and c) from 7th to 18th EO (which is a half of the chain). The straight lines show the distribution for networks without network defects and the dashed lines show the distribution for networks with network defects. The gray dashed line illustrates the distribution of orientations in an ideal PEO melt.}
		\label{angles}
	\end{figure*}
	
	The behaviour of lithium ions in PEO polymer electrolytes was investigated by MD simulations over the last decades \cite{Mao.2000, Maitra.2007, Diddens.2010, Diddens.2014, Polu.2015, Diddens.2017}. In contrast to structures presented in this paper, the PEO chains in earlier studies were not cross-linked. 
	In the following, we will therefore study the polymer orientation and the impact of the cross-linking on the lithium coordination in detail. This might also be important as the orientation can influence the ion transport as stated in the literature. Golodnitsky et al. \cite{Golodnitsky.2002} found in their work that the conductivity is increased in the longitudinal direction along the PEO chains in stretched PEO in comparison to unstretched PEO. In the perpendicular direction, however, the conductivity is decreased as well as the overall conductivity. Therefore, we analysed the network structure for ordering effects, which might influence the ion transport.
	
	Initially, we verified that the network structure is reasonable despite the simplifications that were made. As shown in Figure \ref{structures}e, the lattice collapses during equilibration, resulting in a PEO melt-like structure, fulfilling the requirements for lithium ion transport.
	To verify that our system is equilibrated we splitted the trajectory in several parts and calculated some system properties for our system for each part individually. These properties include the orientation of the polymer chains and the Mean Squared Displacement (MSD), which are both discussed in detail below. We could not observe any significant time dependence within our simulation time for both system properties, concluding that our system is well equilibrated (for further reading see SI).
	
	For the global structural properties it is interesting to understand how the polymer chains are orientated in the equilibrated system. Due to an initially highly ordered underlying network structure we would expect the system to maintain some of its regularity in equilibrated structures at least locally. For very large systems as well as for very long simulation times, we would expect the structure to be isotropic. However, in this work we address only small systems and thus can only observe the local behaviour.  
	
	Therefore, different angles $\Theta$ between the end-to-end vectors of PEO chains and a given spatial direction, shown in Figure \ref{angles}a, were calculated. In a PEO melt of unconnected chains there should be a random distribution \mbox{$p\propto\sin(\Theta)$} of angles, reflecting the fact that for orthogonal orientations more possible realizations can be attained. In the network system, it was observed that the end-to-end vectors of PEO chains are nearly parallel or orthogonal to either the x or y axis or neighboring PEO chains, which are defined as PEO chains that are adjecent within the initial network structure. In fact, this means that the overall structure is still rather ordered. This is supported by the equivalent relative position of PE chains before and after equilibration \ref{structures}. However, from the calculated angles between the vectors connecting non-terminal monomers $n$ and $m$ of a PEO chain and a given spatial direction, it can be seen that there must be local degrees of freedom in the middle of the chain, which is more comparable to PEO melts. Different $n$ and $m$ were chosen as shown in Figure \ref{angles}. For shorter subchains the peaks are lower and flattened. 

	Interestingly, for very short PEO chains the maxima in the distribution are slightly less pronounced and slightly broader, too. It has also be shown that the PEO chains are slightly more coiled in their center than at their ends (see SI, Figure \ref{SI:fig_angles_comparison}).
	
	If network defects are present, the distributions shift in the direction of an unconnected PEO melt, providing even more degrees of freedom (see SI, Figure \ref{angles}) In this sense, networks with defects are an intermediate between polymer melts and ideally formed networks. Among other factors, the exact structure of a system with network defects also depends on the number of defects. 
	
	Considering the length of the end-to-end vectors of PEO chains, there is a much narrower distribution in network systems than in classical PEO melts. The distributions can be fitted by a Gaussian distribution (see Figure \ref{SI:fig_length}).
	For systems with 36 monomers per chain, we get a standard deviation of $\sigma = 4.3 \,\angstrom$ for network systems and $\sigma = 10.4 \,\angstrom$ for non-cross-linked systems. Nevertheless, the average length of end-to-end vectors of polymer chains for network systems is ($R_\mathrm{end}=27.4 \,\angstrom$) only slightly higher than in non-cross-linked systems\linebreak ($R_\mathrm{end}=25.3 \,\angstrom$).
	
	Regarding the local structural properties, one $\mathrm{Li^+}$ is typically coordinated by 4-6 ether oxygens (EO) on average, depending on the employed lithium salt. As observed in previous publications and below the chains wrap helically around the ions. The mean distance of $\mathrm{Li^+}$-EO (2.12\;\AA) is in good agreement with experimental data for this distance \cite{Mao.2000}.
	
	The lithium ions are located mainly in the middle and less at the ends of the PEO chains (see Figure \ref{SI:fig_a}). We expected this to happen due to the absence of EOs in the PE chain, which in conclusion can not coordinate lithium ions. This effect decreases with increasing chain length and can be easily explained by the decreasing ratio of the length of the chain-end region and the length of the whole chain. In the investigated systems the chain-end region, which is irrelevant for ion coordination, consists of four polymer monomers at each end of the chain and is nearly independent of the chain length.
	
	Summing this up, the global structural properties in cross-linked polymers are different than in non-cross-linked polymers. Nevertheless, we observe that the local structural properties remain unaffected by the initial network structure. Lithium ions are mainly coordinated to the center of the PEO chains and only rarely to the chain ends due to the lack of coordinating oxygen atoms at the PE chains.
	
	\subsection{Cation Coordination}	
	\label{cation_coordination}
	
	In the following, we discuss the $\mathrm{Li^+}$ coordination in the network systems in more detail. We calculated distributions of coordination numbers for EOs, PEO chains and anions. 
	
\begin{figure*}[t]
		\centering
		\includegraphics[width=\linewidth]{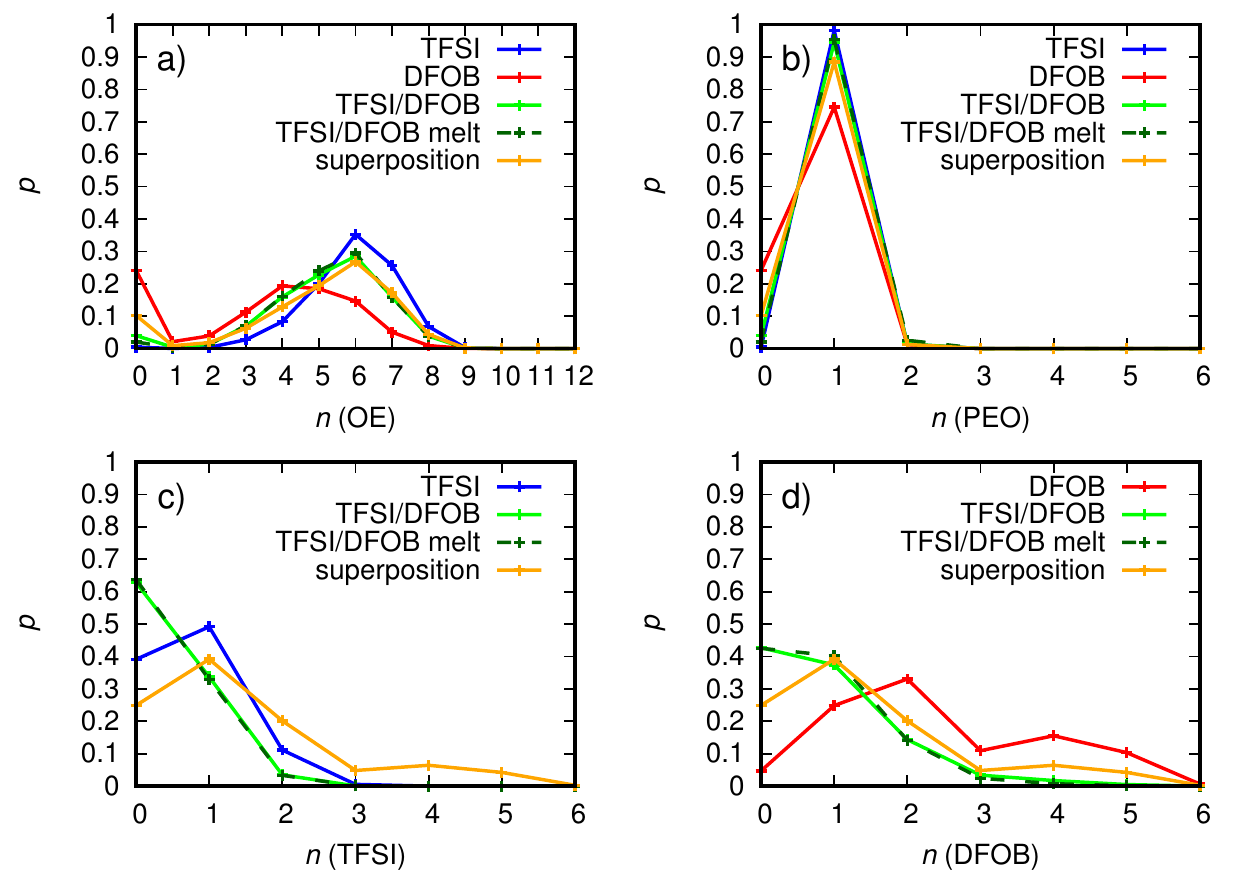}
		
		\caption{The distribution of coordination numbers between lithium  ions and a) ether oxygens, b) PEO chains, c) TFSI anions and d) DFOB anions is shown. For comparison, the distribution is also shown for an amorphous system with dashed lines. The orange line shows the distribution for a dual salt system calculated as a superposition from pure salt systems.}
		\label{coordination}
	\end{figure*}

	First, we will discuss the coordination of $\mathrm{Li^+}$ through PEO ether oxygens and PEO chains. In the following, an EO is considered as coordinated if the Li-EO distance is lower than 4.0\,\AA (see RDFs in SI, \ref{SI:fig_rdf}). For all systems a wide distribution is observed, ranging from two to eight EOs (Figure \ref{coordination}a). The maximum is at six EOs for the TFSI system, which is also the average coordination number. In the DFOB system the distribution is shifted to lower coordination numbers, so that only four EOs are coordinating one $\mathrm{Li^+}$ on average. It is also important to mention that about 20 \% of the $\mathrm{Li^+}$ are not coordinated by ether oxygen atoms in case of LiDFOB, thus rationalizing the larger tendency to form clusters (see below).
	The distribution for the dual salt system essentially corresponds to a superposition of the distributions of the pure salt systems (orange line in Fig. \ref{coordination}). Thus, the average coordination number is five. This superposition only holds for the EOs, but can not reproduce the distributions discussed below.
	
	A $\mathrm{Li^+}$ is considered as coordinated to a specific PEO chain if at least one ether oxygen of this chain coordinates the lithium ion. In systems with TFSI, more than 99.5\,\% of all $\mathrm{Li^+}$ are coordinated by at least one PEO chain. For DFOB this number is clearly lower around 75\,\% to 85\,\% depending on system size and time scale (see below). For the dual salt it was observed that more than 97\,\% of lithium ions are coordinated at least through one PEO chain. Lithium ions that are coordinated through two chains are observed very rarely (see Figure \ref{coordination}b). 
	
	As it would be expected, we found no significant effect of the PEO chain length on the distributions of coordination numbers (see Figure \ref{SI:fig_coordination}). They are only dependent on the employed lithium salt at a given temperature. A technical challenge that emerged during calculation is clustering of $\mathrm{Li^+}$ and DFOB. Due to large and stable clusters, even though they highly fluctuate in size, sometimes a significant variance in the distribution of PEO occurs, which we discuss in detail below. Nonetheless, we emphasize that the distributions in Figure \ref{coordination}b are representative for the systems with 24 monomers per chain. 

	Second, we checked how anions are coordinating to lithium ions (Figure \ref{coordination}c and d). Our analysis revealed that TFSI mainly coordinates $\mathrm{Li^+}$ through oxygen atoms (considered as coordinated if the distance between oxygen and $\mathrm{Li^+}$ is lower than 3.3\;\AA). For DFOB we observed a significant coordination through carbonyl oxygen atoms (considered as coordinated if the distance is lower than 3.3\;\AA) and fluorine atoms (considered as coordinated if the distance is lower than 3.0\;\AA). Other atoms coordinate $\mathrm{Li^+}$ only rarely.
	We have observed that DFOB has a higher affinity to coordinate $\mathrm{Li^+}$ than TFSI. Whereas $\mathrm{Li^+}$ is mostly coordinated through no or one TFSI, it is often coordinated through more than two DFOB in the respective pure salt systems. In the pure salt systems, 40\% of $\mathrm{Li^+}$ are not coordinated through any TFSI and only 5\% of $\mathrm{Li^+}$ are not coordinated through any DFOB. The average coordination number is lower than one for pure TFSI and approximately two for pure DFOB. In comparison to pure DFOB systems the coordination number of DFOB is decreased dramatically if fewer DFOB molecules are present and replaced by TFSI molecules in the dual salt systems.
	In total, the coordination nubmers are similar to our preliminary results from Shaji et al. \cite{Shaji.2022}, in which the network was modeled as a PEO melt with fixed chain ends, demonstrating that the local lithium coordination is insensitive to the precise network structure.
		
	\subsection{Cluster}
	
	\begin{figure*}[t]
		\centering
		\includegraphics[width=\textwidth]{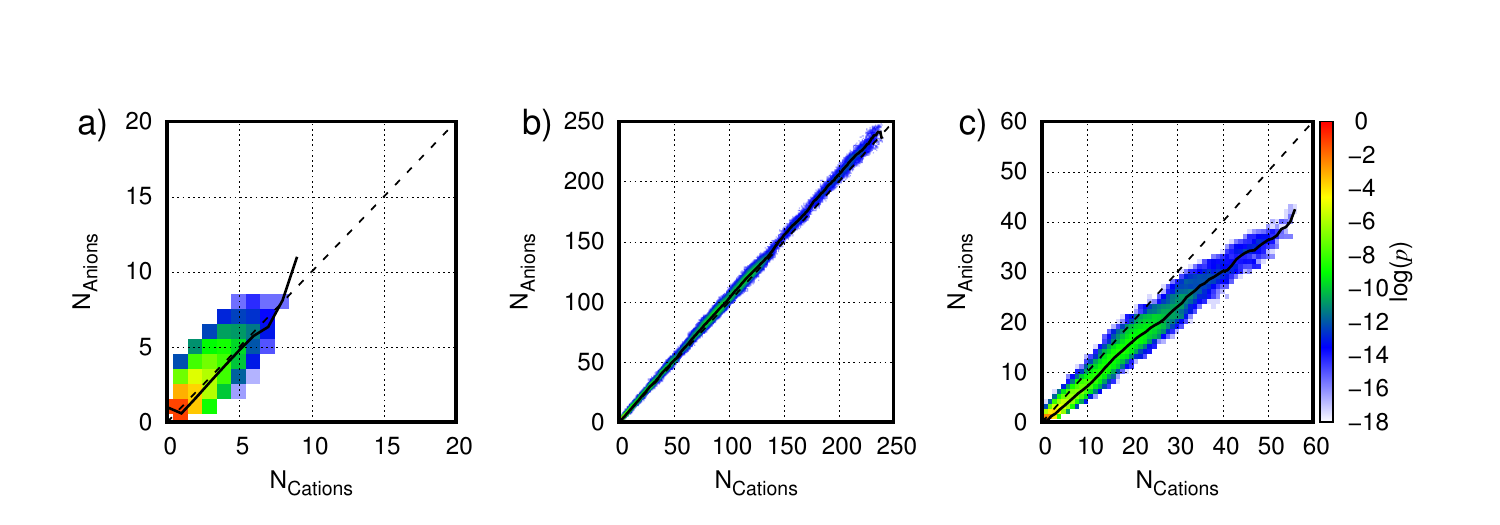}
		
		\caption{Logarithmic probability to find a cluster with a certain number of cations $N_\mathrm{Cations}$ and anions $N_\mathrm{Anions}$ at any time of the simulation. The composition of the clusters in the reference systems for a) pure LiTFSI salt system, b) pure LiDFOB salt system and c) dual salt system is shown. The black dashed lines indicate the diagonal and the straight black lines represent the average number of anions in a cluster as a function of the number of cations.}
		\label{heatmap}
	\end{figure*}

	Figure \ref{coordination} contains information about the coordination of a given lithium ion. Next, we study the distribution of ion clusters involving multiple ions. It was observed that TFSI is well distributed in the system. It coordinates mainly with one oxygen atom to $\mathrm{Li^+}$ (88\,\%), but it is also able to coordinate with two oxygen atoms to one $\mathrm{Li^+}$.
	The distribution of DFOB in the system is more complex. From the snapshot in Figure \ref{SI:fig_cluster_snapshot} it is easy to see that DFOB and $\mathrm{Li^+}$ are clustering, as also observed by visual inspection in our previous analysis \cite{Shaji.2022}. It was observed that DFOB mainly coordinates only with one atom to one $\mathrm{Li^+}$ (95\%), but in contrast to TFSI it coordinates often two or three different $\mathrm{Li^+}$. In addition, one $\mathrm{Li^+}$ is often coordinated by two or more DFOB. This enables DFOB and $\mathrm{Li^+}$ to form clusters of a size that can be as large as the system size. As a consequence, the cluster sizes observed in the simulations only provide a lower bound when estimating the true cluster size. For the same reason, the exact $\mathrm{Li^+}$-DFOB coordination numbers and the distribution of coordination numbers for PEO and lithium ions may slightly differ (see Section \ref{cation_coordination}). Furthermore, it is important to mention that the clusters are stable over long time scales. The effect of clustering on the systems dynamics will be discussed later.  
	
	Nonetheless, we quantify the clustering behaviour in the following, thereby systematically extending our previous work \cite{Shaji.2022}. Because we observe huge clusters in DFOB systems, we used the larger reference systems with a shorter simulation time for this purpose (see Section \ref{sec_MD_simulations}). We defined a cluster of consisting of at least one anion and one cation. In a cluster all ions are connected directly or through other ions to each other. 
	Note that due to this definition it might be helpful to think of the cluster as a network structure instead of a spherical shape.
	
	The heatmaps in Figure \ref{heatmap}a-c show the logarithmic probability to find a cluster of a defined size at a certain moment during the simulation. We observe that TFSI forms very small clusters with lithium ions that are smaller than 10 anions and cations each (Figure \ref{heatmap}a). Unfortunately, the clusters DFOB forms with lithium ions are still very huge in comparison to the total system size (Figure \ref{heatmap}b). They can consist of more than 240 anions and cations each at their maximum size. Thus, even larger clusters in larger systems could be possible. Nevertheless, most clusters are smaller than 150 anions and cations each. We observe a bimodal distribution with one maximum in the region of very small clusters with less than 10 cations and anions each and another maximum in the region of clusters between 50 and 150 anions and cations each.
	For the dual salt system (Figure \ref{heatmap}c) we observe smaller clusters than in the DFOB system and larger clusters than in the TFSI system. The maximum size is lower than 60 cations and 50 anions. Interestingly, the dual salt system tends to form clusters, which tend to consist of more cations than anions, whereas in the pure salt systems clusters consist of a nearly equal numbers of cations and anions.
	Molinari et al. observed for various pure salts highly negatively charged clusters in ionic liquids \cite{Molinari.2019}, whereas they are almost neutral in our study. This might be an effect of the polymer and is supported by further results from Molinari et al.\cite{Molinari.2018} who found neutral clusters for low salt concentrations and negatively charged clusters for high salt concentrations in polymer melts. Ion clusters in a dual salt system, for which we found positively charged clusters, was not studied in literature. 
	
	In the work of Molinari et al., it was assumed that as a consequence of an excess negative charge of these clusters they are moving in the opposite direction than isolated lithium ions during battery operation, reducing the transference number. Similarly, we assume that as a consequence of an excess of positive charge the clusters might move in the same direction as isolated lithium ions during battery operation, increasing the transference number, although the underlying mechanism is likely more complex and one cannot quantitatively derive dynamical information from structural properties \cite{Wettstein.2022}. 
	
	We speculate that clusters with an excess of lithium ions are caused by the different ability of TFSI and DFOB to coordinate lithium ions. As visual inspection shows, larger clusters mainly consist of DFOB and $\mathrm{Li^+}$ in the core. TFSI attaches mostly at in the terminal region of the clusters. This means  that the clusters are only rarely terminated by DFOB, but often by TFSI or $\mathrm{Li^+}$. Whereas the core is more stable, TFSI can attach and detach comparably fast (see mean residence times, Section \ref{Sec:residence_times}). To explain the positive charge in a simple picture, assume a neutral cluster with this structure. Once TFSI detaches the cluster becomes positive and there is an uncoordinated TFSI. To become neutral again, a lithium ion, which is more probably coordinated by DFOB, must detach, which is unlikely, or the TFSI must come back. Unless a TFSI ion enters the cluster again, it has a positive net charge. The free $\mathrm{Li^+}$ coordination site in turn can also be saturated by PEO (Figure \ref{coordination}). If there are many terminating TFSI which detach, the cluster has a larger positive net charge.
	Nevertheless, further investigations would be needed to understand the effect and process of clustering in detail in the dual salt system. 
	
	Moreover, we analysed the largest cluster in the pure DFOB systems to qualitatively determine the fluctuations in size and stability of clusters. We found out that the largest cluster is stable during the complete simulation time, which might be a finite size effect since the cluster can be stabilized by the periodic boundary conditions. Its size fluctuates mainly between approximately 200 and 250 anions and cations in total. Time periods up to approximately 25\;ns were observed, in which the cluster is stable at this size (see Figure \ref{SI:fig_cluster_size}). On larger time scales the size of the cluster fluctuates even more due to merging with another cluster or separation of certain parts of the cluster. Thus, the cluster can exceed 400 anions and cations in total. Sizes below 100 anions and cations are observed very rarely for the largest cluster.
	To quantify the fluctuation of the size of the largest cluster, the cluster size was averaged over 1 ns and then the coefficient of variation  $c_v$ was calculated (for further reading see SI). We receive values for $c_v$ mainly between 0.05 and 0.4. Note that $c_v$ is low for time periods, in which the cluster does not interact with other clusters.
		
	\subsection{Cation Dynamics}
	\subsubsection{Residence Times}
	\label{Sec:residence_times}
\begin{figure*}[t]
		\centering
		\includegraphics[width=\linewidth]{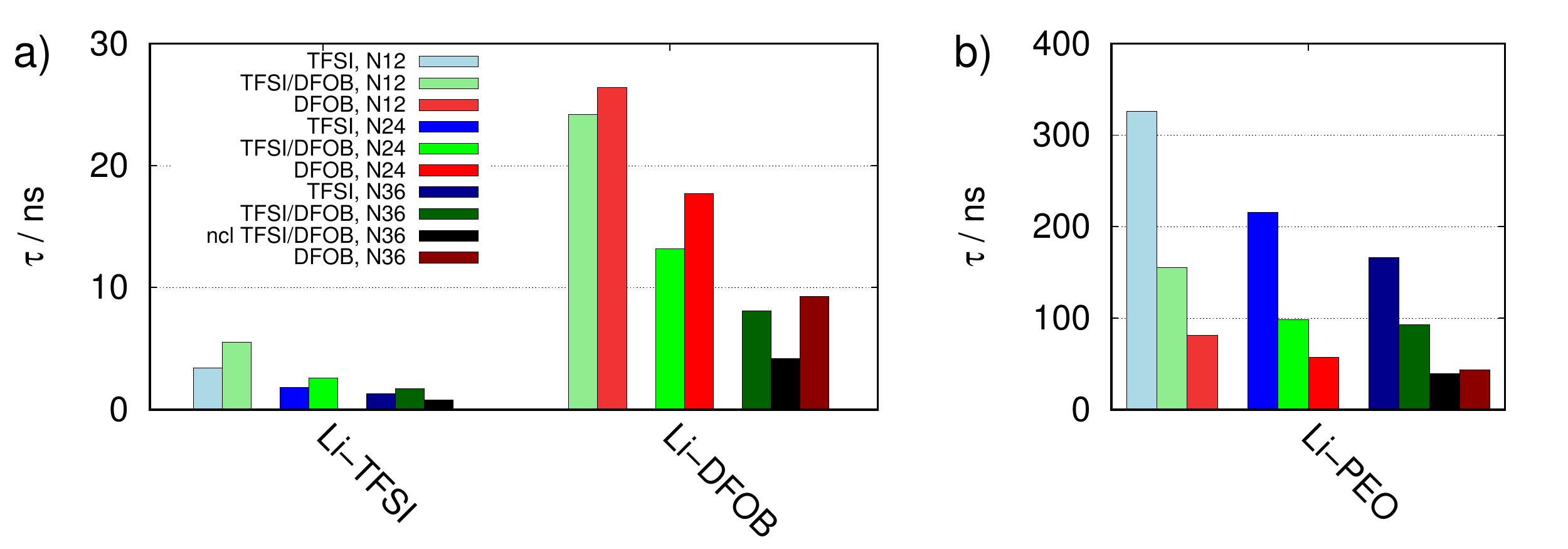}
		
		\caption{The mean residence time $\tau_\mathrm{r}$ is shown for a) TFSI and DFOB and b) PEO. Systems with different PEO chain lengths are compared. For the dual salt system with a PEO chain length of 36 also values for non-cross-linked (ncl) polymers are given.}
		\label{res_time}
	\end{figure*}

	Next, we analysed the cation dynamics in order to understand the transport mechanics of ${\mathrm{Li}^+}$ in network systems. 
	The first important physical value we consider is the mean residence time of ${\mathrm{Li}^+}$, which can be interpreted as the average time $\tau_r$ that a lithium ion coordinates to either PEO or anions before it jumps to another chain or ion cluster. 
	To calculate the mean residence time, we first define the time correlation function as reported in the literature \cite{Zhao.2009} 	
	\begin{equation}
		\begin{split}
			C(t) & = \frac{\langle \delta h(0) \delta h(t)\rangle}{\langle \delta h^2 \rangle} = \frac{\langle h(0)h(t)\rangle - \langle h \rangle ^2}{\langle h \rangle - \langle h \rangle ^2} \\
			& = \frac{\langle h(0)h(t)\rangle}{\langle h \rangle}.
		\end{split}
	\end{equation}  
	In this calculation, $h(t)$ is one if ${\mathrm{Li}^+}$ is coordinating either to a given PEO chain for the ${\mathrm{Li}^+}$-PEO mean residence time or an anion for the ${\mathrm{Li}^+}$-anion mean residence time and zero if ${\mathrm{Li}^+}$ is not coordinating. $\langle h \rangle$ is the average number of PEO or anions coordinating a ${\mathrm{Li}^+}$. In fact, we checked for a ${\mathrm{Li}^+}$ that is coordinating at time $t_0$ to a specific PEO or anion, if it is coordinating still at time $t_0+t$. We averaged over all starting times $t_0$ in our simulation. Note that it is irrelevant for this calculation what happens to the specific ion between $t_0$ and $t_0+t$, therefore also ions that detach and jump back to their initial coordination partner are captured. 
	The mean residence time can be calculated by the integral of the time-correlation function in a second step. To integrate the resulting decay curve $C(t)$, we applied a fit to the time-correlation-function, which has the form of a Kohlrausch-Williams-Watts-function as reported in literature \cite{Zhao.2009}
	\begin{equation}
		c(t) = \exp{\left( - \left(\frac{t}{\tau}\right)^{\beta} \right)}
	\end{equation}
	and then calculated the integral of it as
	\begin{equation}
		\tau_\mathrm{r} = \frac{\tau}{\beta} \Gamma \left(\frac{1}{\beta}\right).
	\end{equation}
	$\beta$ and $\tau$ are fit parameters, whereas $\tau_\mathrm{r}$ is denoted as the mean residence time, which we identify as $\tau_3$ in the case of the ${\mathrm{Li}^+}$-PEO mean residence time. $\Gamma$ is the gamma function.
	We calculated the mean residence time for ${\mathrm{Li}^+}$ with PEO, TFSI and DFOB. The results are shown in Figure \ref{res_time}.
	
	Comparing the results for different systems, we are able to derive two main results. First, in general, $\mathrm{Li^+}$ ions are coordinating much longer to one specific PEO chain than to a specific anion. Comparing the anions TFSI and DFOB, $\mathrm{Li^+}$ coordinates only briefly in a time scale of a few nanoseconds to TFSI, but up to 30 ns to DFOB. In agreement with our previous work \cite{Shaji.2022}, we observe that the PEO residence time is significantly larger for the TFSI system than for the DFOB system, with the dual-salt system showing an intermediate value. Second, the mean residence time in general depends on the chain length of PEO. For a given system it decreases for both PEO and anions when the chain length is increased.
	Going further into detail, we identify the system with TFSI and 12 monomers per chain as the system with the longest mean relaxation time for PEO with $\tau_\mathrm{3}=326 \mathrm{\, ns}$. In contrast, for a system with 36 monomers per chain and DFOB we observed $\tau_\mathrm{3}=43 \mathrm{\, ns}$. Referring to the transport model, we would expect a short mean residence time to be optimal for effective ion transport. 
	Note that the mean residence time for PEO calculated for systems with non-cross-linked PEO melts is significantly lower. For example, we observe $\tau_3 = 48\;\mathrm{ns}$ in the non-cross-linked system whereas we observe $\tau_3 = 92\;\mathrm{ns}$ in the cross-linked dual salt system with 36 monomers per PEO chain.
	For systems with network defects, we observe mean residence times that are in between those for ideal cross-linked networks and non-cross-linked networks (see Table \ref{results}). 	
	
	Together with the broadened distribution of angles mentioned above this indicates that the orientation of PEO chains may play an important role for the jumping process, as longer chains in the cross-linked systems are more flexible and thus allow a larger degree of possible local orientation of two PEO chains relative to each other, which is considered to be important for the success of a jumping process. This would also rationalize why the mean residence times in systems with defects or even melts are generally lower as compared to fully cross-linked chains.
		
	Moreover, this means that also the mobility of PEO chains, which must be higher in systems with network defects or PEO melts, has an impact on the time, which an ion needs to leave a PEO chain and jump to another one. Note that in earlier publications the mean residence time was assumed to be independent from the mobility of PEO chains \cite{}. However, this is only valid if the local intermolecular environment of a $\mathrm{Li^+}$ coordinated to PEO relaxes on a similar time scale for all chain lengths, which should be fulfilled for all but very short chains. For networks, the relaxation of the local surroundings of a $\mathrm{PEO-Li^+}$ complex is strongly affected by the cross-linking, such that it also influences the mean residence time. 
	
	\subsubsection{Motion along the backbone}
	\label{Sec:motion_along_backbone}

	\begin{figure*}[t]
		\centering
		\includegraphics[width=\linewidth]{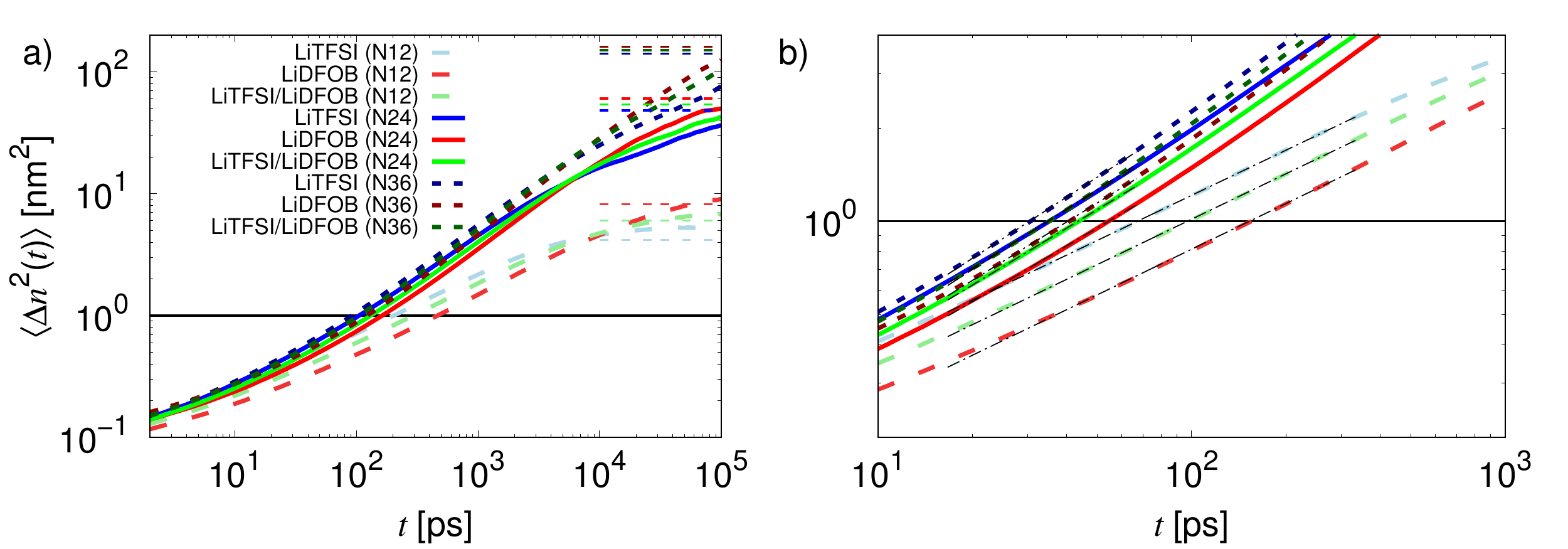}
		
		\caption{The one dimensional MSD $\langle \Delta n^2(t)\rangle$ of lithium ions along the polymer backbone is shown for different systems. The black dashed lines are the fits used to calculate $\tau_\mathrm{1,eff}$. The colored dashed lines show the theoretically expected plateau value of the MSD for each of the systems in the same color, using the reduced chain length $(N-a)$. a) shows the whole plot, b) only the region used for the fit.}
		\label{n_MSD}
	\end{figure*}	
	
	To elucidate how long an ion needs to explore a whole chain we characterized the motion of Li along the PEO chain. Typically 2 to 8 neighbouring ether oxygens of the backbone are coordinating to one $\mathrm{Li^+}$ (Figure \ref{coordination}). With evolving time, the $\mathrm{Li^+}$ ion can explore the total chain, so that the coordinating ether oxygens within a chain change. We assign an index to each EO atom to identify it clearly. Then we determine an average monomer index $n$, to which $\mathrm{Li^+}$ is coordinating at a given time. The one dimensional mean squared displacement (MSD) ($\langle \Delta n^2(t) \rangle$) along the index $n$ is shown in Figure \ref{n_MSD}. Furthermore, we show the theoretically expected plateau values of the MSD, which can be simply calculated as
	\begin{equation}
	\mathrm{Plateau} = \frac{(N-a)^2}{6}
	\end{equation}
	using a reduced chain length $(N-a)$. The calculated plateau values are in good agreement with the observed plateau values. The estimation of $a$ is described in the SI (Section \ref{SI:sec_a}).

	In general, it is possible to calculate a diffusion coefficient by using the Einstein equation 
	
	\begin{equation}
		D = \lim_{t\to\infty} \frac{\langle\Delta n^2(t) \rangle}{2t}
		\label{Eq_einstein}
	\end{equation}
	from which the relaxation time $\tau_1$ may be obtained.
	Unfortunately, we observe subdiffusive behaviour with \linebreak $\langle \Delta n^2(t) \rangle \propto t^{0.3} \textrm{ to } t^{0.8}$ depending on the exact system and time scale. Therefore the dynamics is still slightly non-Markovian, so that we cannot calculate the diffusion coefficient directly. Furthermore, the MSD reaches a plateau value during $\tau_\mathrm{3}$, indicating that on average the ions have already fully explored the coordinating PEO chain when a transfer occurs. 
	
	In this context, note that ideally the above discussed mean residence time should be at least shorter than the time the ion needs to explore the total PEO chain. This is due to the fact that after exploring the PEO chain the ion can only be additionally transported by the polymer dynamics unless it is transferred to another chain. As we see later and as it would be expected due to cross-linking, the center of mass movement of PEO chains is negligible and thus ineffective for ion transport in cross-linked systems.
	
	To nonetheless determine and compare the behaviour, we define an effective time $\tau_\mathrm{1,eff}$, which is the time where $\langle \Delta n^2(t)\rangle=1$. We calculated $\tau_\mathrm{1,eff}$ using a fit of the type 
	\begin{equation}
 		\langle\Delta n^2(t) \rangle = c\cdot t^{\alpha}
 	\end{equation}
 	with the fitting parameters $c$ and $\alpha$. In the picture of the transport model $\tau_\mathrm{1,eff}$ is the time a lithium ion needs to move from one ether oxygen to the next along a PEO chain. This time is in the order of picoseconds and decreases with increasing chain length and is also dependent on the chosen salt \ref{results}. It is shorter for the TFSI systems and longer for DFOB systems and intermediate for the dual salt systems. The chain length dependency can be explained by the fact that in shorter chains the lithium ion can only move very restricted along the chain due to the inaccessible chain end. The salt dependency can be explained by the stronger coordination of DFOB to $\mathrm{Li^+}$ and clustering, which can also slow down and restrict the ion transport along a chain, although the lithium ions are less strongly bound to PEO in DFOB systems due to the lower coordination number of ether oxygen atoms, from which a faster movement along the chain would be expected. The coordination of lithium ions through DFOB has therefore a larger impact on this specific dynamics than the coordination through ether oxygen atoms of PEO. Furthermore, lithium ions that are coordinated by PEO and the coordinating ether oxygen atoms are also slower in the DFOB systems than in the TFSI systems (see Figure \ref{SI:fig_msd_polymer_def}), supporting the above explanation.
 	This indicates that the ion transport along the chain is more efficient in the TFSI systems, whereas the ion transport from one PEO chain to another through jumping is more efficient in the DFOB systems.
 	
 	In cross-linked systems with network defects $\tau_\mathrm{1,eff}$ is only very marginally decreased in comparison to ideally cross-linked systems. In non-cross-linked systems it is further decreased. We observe $\tau_\mathrm{1,eff}=80\;\mathrm{ps}$ in the non-cross-linked system and $\tau_\mathrm{1,eff}=104\;\mathrm{ps}$ in the cross-linked system (both with TFSI and DFOB and 36 monomers per chain).
 	
 	\subsubsection{Mean Squared Displacement}
 	 	
\begin{figure}[t]
 		\centering
 		\includegraphics[width=\linewidth]{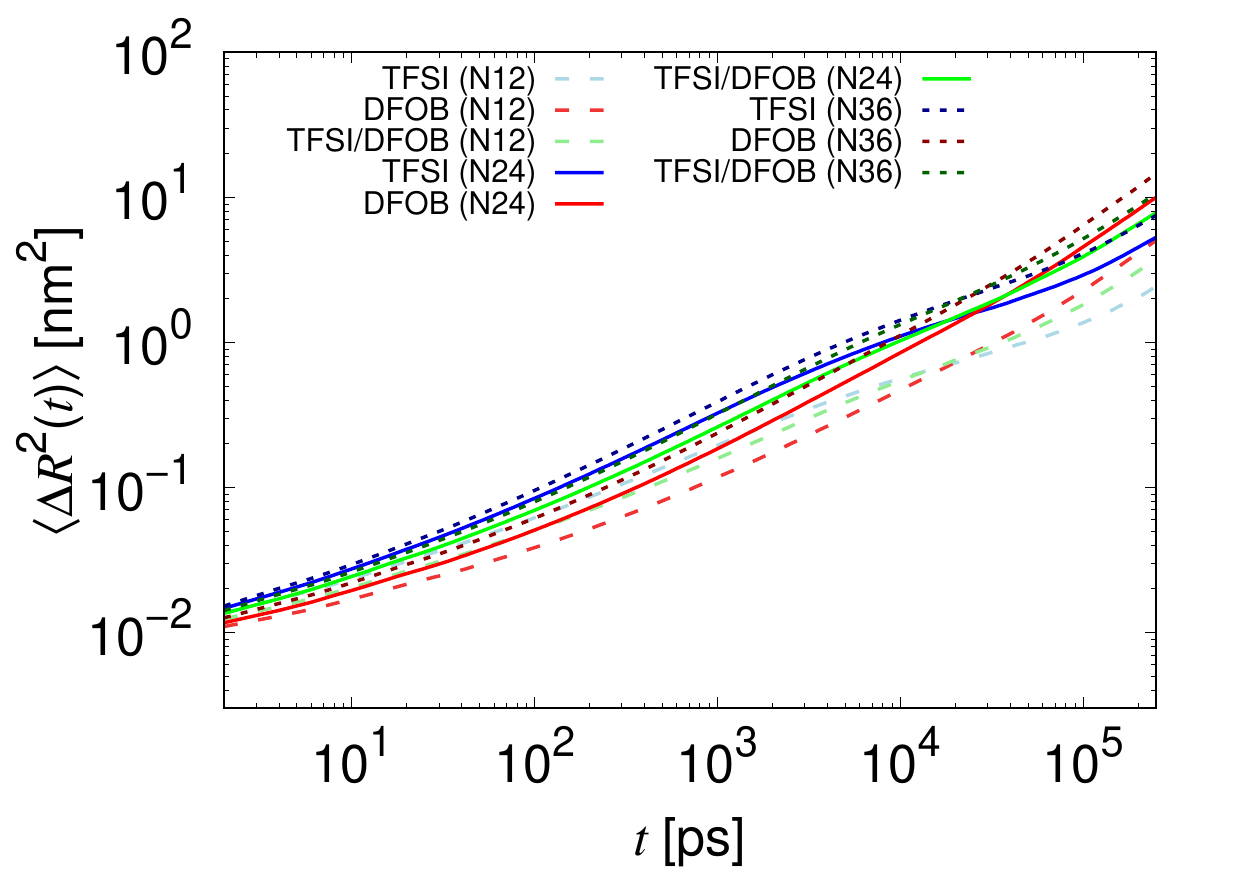}
 		
 		\caption{MSD of lithium ions in systems with 12, 24 and 36 PEO monomers per chain.}
 		\label{total_msd}
 	\end{figure}

 	Next, keeping the above findings in mind, we will discuss the total MSD of lithium ions. From Figure \ref{total_msd}, which shows the total MSD for $\mathrm{Li}^+$ in different systems, we qualitatively observe the same trend for each chain length. In general, the MSD increases with increasing chain length. As observed in the previous 1-dimensional MSD along the PEO chains, the MSD in systems with TFSI is higher for short time scales and the MSD in systems with DFOB is higher for long time scales. 
 	The effect of increasing the chain length from 12 monomers per chain to 24 monomers per chain is for example comparable to the effect of using LiDFOB instead of LiTFSI.
 	We observe that the slope of the MSD in DFOB systems increases continuously, becoming more diffusive with evolving time. Interestingly, in TFSI systems, the slope increases first for short time scales of a few nanoseconds and then decreases before increasing again. This leads to an intersection of the MSDs for different salts, which is dependent on chain length, between 20 ns and 30 ns, at which the dynamics of $\mathrm{Li}^+$ in DFOB systems becomes faster than in TFSI systems. As can be concluded from the mean residence times $\tau_r$ and also has been revealed by a more detailed analysis jumping of $\mathrm{Li}^+$ from one chain to another or to anions and vice versa (see SI, Section \ref{SI:sec_jumping}) this trend can be explained as follows: The mean residence time of a $\mathrm{Li}^+$ at a PEO chain is highly affected by the presence of anions. Jumps from one PEO chain to another with no anions involved are observed very rarely. In systems with DFOB we observe a mean residence time ($\tau_3 < 82\,\mathrm{ns}$) that is drastically lower than in TFSI systems ($\tau_3 > 166\,\mathrm{ns}$, see Table \ref{results}). This is caused by the higher affinity of DFOB to coordinate $\mathrm{Li}^+$ and by a longer mean residence time $\tau_r^\mathrm{Li^+/Anion}$ in salt clusters of DFOB systems. This simply gives DFOB molecules more time to support $\mathrm{Li}^+$ during the jumping process. 
 	
 	Consequently, also the number of jumps between PEO chains and anions is increased dramatically in systems with DFOB in comparison with systems with TFSI. 
 	To analyse the MSDs quantitatively, we fitted functions of the type
 	\begin{equation}
 		\langle\Delta R^2(t) \rangle = c\cdot t^{\alpha}
 		\label{msd_fit}
 	\end{equation}
 	to the data in three different ranges. The exponent $\alpha$ indicates the diffusive behaviour of the lithium ions. In the range of 0.1 to 1.0 ns we observe $\alpha \approx 0.50 \textrm{ to } 0.60$. 
 	This is roughly compatible with the subdiffusive motion expected from the Rouse model with $\alpha = 0.5$, indicating the cooperative motion of the lithium ions with the polymer segments (see also below). 
 	The diffusive behaviour depends on the chain length, $\alpha$ increasing with $N$, but is independent from the lithium salt. 
 	Next we applied a fit in a range of 1.0 to 10 ns, expecting that this would be a time scale, at which salt effects become significant, due to previous results. We observe $\alpha \approx 0.45 \textrm{ to } 0.55$ for TFSI systems. As described above the motion of lithium becomes more subdiffusive because the lithium ions cannot move further while coordinated to the same polymer chain (see Section \ref{Sec:motion_along_backbone}). In contrast, we observe $\alpha \approx 0.60 \textrm{ to } 0.70$ for DFOB systems meaning that the motion of lithium ions becomes more diffusive. The exponents in the dual salt system are comparable to the exponents at shorter time scale. 
 	This indicates that DFOB accelerates lithium ion dynamics with evolving time, whereas TFSI slows it down for long time scales. However, the motion of lithium ions is faster on time scales of 1-10\;ns in TFSI systems than in DFOB systems, as can be seen from Figure \ref{total_msd}. In a dual salt system these two effects cancel out at this time scale.
 	Moreover, we applied a fit in the range of 150 to 250 ns to study systems dynamics for comparatively large time scales. Due to the logarithmic scale this fit is only a very rough estimate, but even larger time scales are not available via MD simulations. Unfortunately, we observed still subdiffusive behaviour for all network systems. Interestingly, the diffusive behaviour is more dependent on the chosen lithium salt, than on the chain length. We observe $\alpha \approx 0.7$ for TFSI systems, $\alpha \approx 0.8$ for DFOB systems and $\alpha \approx 0.9$ for dual salt systems. That $\alpha$ is largest for the dual salt systems may be a hint that on larger time scales it may perform better than both pure salt systems, as observed in the experiments \cite{Shaji.2022}. 
 	
 	We also studied the dependence of lithium dynamics in relation to different spatial directions. Because of the initial order of the network systems one might expect that the dynamic in z-direction of the networks may be different than in x- and y-direction. A calculation of 1D-MSDs in x-,y- and z-direction has shown that a significant difference is only observable for TFSI and dual salt systems. The MSD in z-direction is lower than in x- and y-direction for TFSI systems, whereas for the DFOB systems the effect is marginal (see SI, Figure \ref{SI:fig_msd_direction}).
 	In general, the difference is only observed for time scales larger than approximately 10 ns. The explanation for this observation is that at short time scales lithium ions are able to move move cooperatively with the chain. With increasing time lithium ions need to jump to other PEO chains or anions (in z-direction) to explore more space. As described already, anions play an important role in jumping process. Because in TFSI systems we only observed few jumps, movement in z-direction is slower in these systems. In DFOB systems there are sufficient jumps to enable lithium to move in each direction equally fast. 
 	
 	Finally, note also that there are different barriers for ion motion within the clusters for LiTFSI and for LiDFOB. One might speculate that this might make the systems dynamics highly temperature dependent. Salts that are leaning towards clustering, have higher barriers for ion motion \cite{Nair.2019}. Following this argumentation, it might be interesting to determine mean residence times and MSDs for ions inside and outside clusters separately to gain further insights and to explain deviations between experiment and simulation, which were carried out at different temperatures.

	\subsection{Polymer Dynamics}
\begin{figure*}[t]
		\centering
		\includegraphics[width=\linewidth]{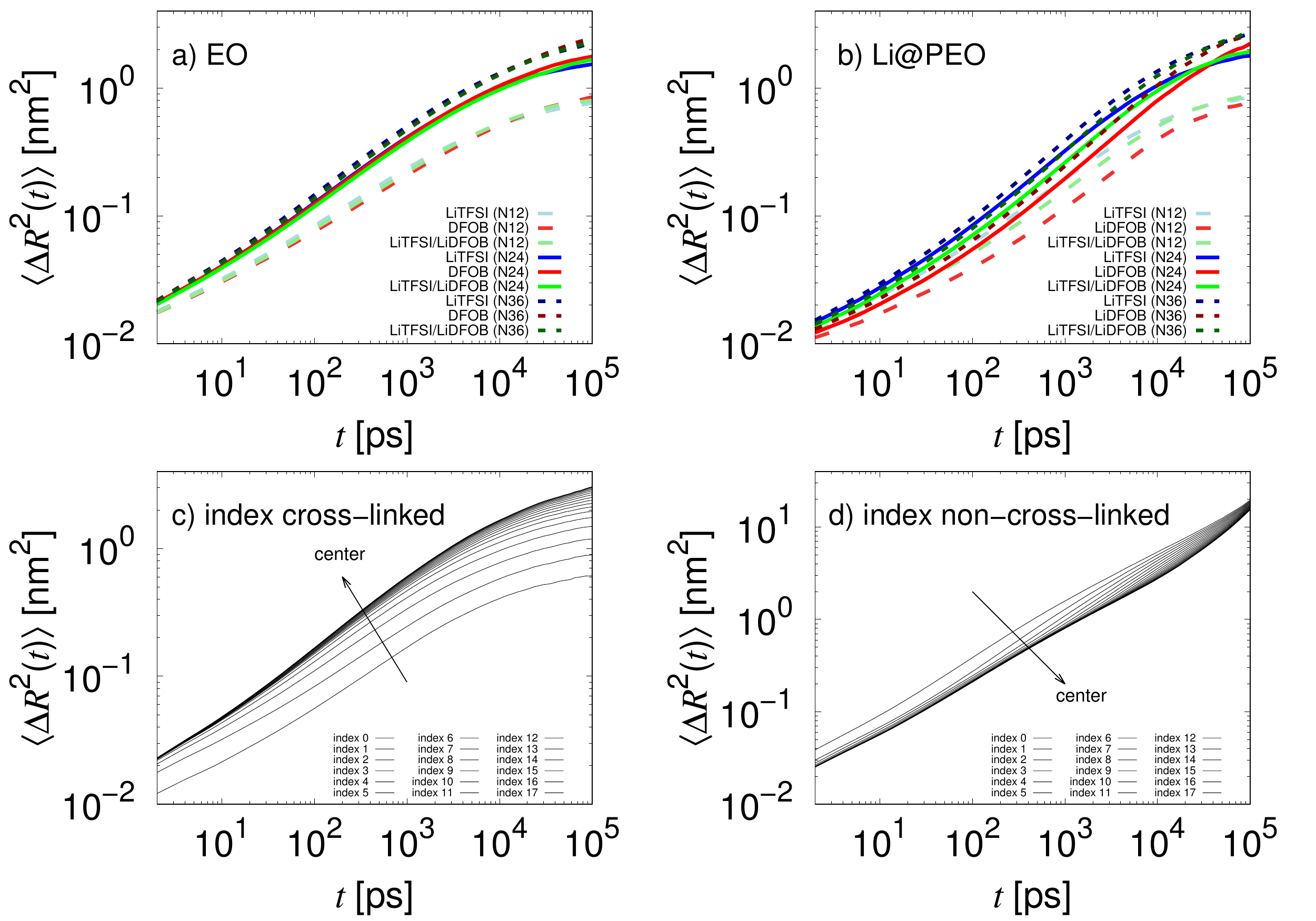}
		
		\caption{The average MSDs $\langle\Delta R_\mathrm{EO}^2(t)\rangle$ of the ether oxygens of PEO in a cross-linked polymer as a function of time $t$ are shown in a). The MSD $\langle\Delta R_\mathrm{Li@PEO}^2(t)\rangle$ of lithium ions that are bound to a specific PEO chain for at least time $t$ are shown in b). In c) the MSD of the ether oxygen atoms is shown with respect to their index, which is 0 for the first group in a polymer and $N-1$ for the last group, where $N$ is the number of monomers. The MSD of ether oxygens in the center of a chain is larger than at the ends. Similar is shown in d) for non-cross-linked polymers.}
		\label{msd_polymer}
	\end{figure*}

	In addition to the MSDs of lithium ions, we also calculated the MSDs of ether oxygens and the MSDs of lithium ions, which are bound to a specific PEO chain for at least time $t$. The additional data enabled us to estimate the impact of the polymer mobility on the lithium dynamics (i.e. the second mechanism of the lithium ion transport model).
	The MSD $\langle\Delta R_\mathrm{EO}^2(t)\rangle$ of the ether oxygens is shown in Figure \ref{msd_polymer}a. It depends mainly on the length of the PEO chains and only weakly on the lithium salt. With increasing chain length the MSD increases. $\langle\Delta R_\mathrm{EO}^2(t)\rangle \propto t^{0.5}$ is observed for all systems between 0.1 and 10 ns, as it would be expected from the Rouse model \cite{Rouse.1953,Rouse.1998}. For larger time scales the motion becomes more subdiffusive as a result of fixed ends of the PEO chains.
	Figure \ref{msd_polymer}c shows as an example the average MSD of individual ether groups according to their index, which we described previously, in cross-linked systems. We observe that ether oxygens at the end of the PEO chains are significantly slower than in the center. This is opposite to the observation in non-cross-linked systems, shown in Figure \ref{msd_polymer}d, in which the ether oxygens at the end are slightly faster than in the center at least for short time scales. For larger time scales when the center of mass movement becomes dominating the MSD becomes independent of the group index. 
	
	Additionally, in Figure \ref{msd_polymer}b the MSD $\langle\Delta R_\mathrm{Li@EO}^2(t)\rangle$ of the lithium ions that are bound to a specific PEO chain is shown, which we need to estimate diffusion coefficients in a next step.
	For the calculation of $\langle\Delta R_\mathrm{Li@PEO}^2(t)\rangle$ short jumps are ignored, if the lithium ion jumps back to the PEO chain after a few picoseconds (for further reading on the method see supporting information).  
	On short time scales, the MSD of lithium ions bound to a PEO chain is slightly lower than the MSD of the PEO chains due to the additional degrees of freedom of the polymers. For larger time scales, it is slightly larger due to the possible movement of the ions along the chain. Nonetheless $\langle\Delta R_\mathrm{Li@PEO}^2(t)\rangle$ roughly follows $\langle \Delta R^2_{EO} \rangle$, indicating the cooperative motion of ions with the polymer chains.
		
	\subsection{Diffusion Coefficients and ion transport number}
	
	Finally, we complete our analysis of the transport mechanism by estimating the overall lithium ion diffusion coefficient and the transference number $t_{Li}$ using the previous results.
	
	It is not possible with the available data to calculate the lithium ion diffusion coefficients directly according to the Einstein relation due to the subdiffusive behaviour. In the following, we outline how we use the available data to estimate the diffusion coefficients nevertheless.
	To obtain $D_\mathrm{Li}^\mathrm{PEO}$ and $D_\mathrm{Li}^\mathrm{anion}$ we calculated the MSDs of lithium ions that are bound to a specific PEO chain or a specific anion at least for time $t$ as shown in Figure \ref{msd_polymer} (and Figure \ref{SI:fig_msd_Li@anion} for lithium ions bound to anions).
	
	On the lines of the previous calculation, the diffusion coefficients $D_\mathrm{Li}^\mathrm{PEO}$ and $D_\mathrm{Li}^\mathrm{anion}$ can be obtained approximately as
	\begin{equation}
		\label{Eq_diffusion_coefficient}
		D_\mathrm{Li}^\mathrm{species} = \frac{\langle\Delta R_\mathrm{Li}^2(\tau_\mathrm{r})\rangle}{6\tau_\mathrm{r}}
	\end{equation} 
	due to the fact that the jump process of the lithium ions - either to another PEO chain or to the anions or vice versa - can be interpreted as renewal events, such that the dynamics becomes Markovian after $\tau_\mathrm{r}$. This allows us to estimate $D_\mathrm{Li}$ despite the subdiffusive MSDs in Figure \ref{total_msd}. Note that Equation \ref{Eq_diffusion_coefficient} was used for both lithium ions coordinated to PEO as well as to anions.
	
	Because the total lithium ion MSD is subdiffusive, we estimated the total diffusion coefficient $D_\mathrm{Li}^\mathrm{tot}$ as
	\vspace{-5cm}
	\begin{table*}[ht]
		\caption{Percentage $p_\mathrm{PEO}$ of lithium ions that are bound at least to one PEO chain, relaxation times $\tau_1$ and $\tau_3$ defined by the lithium ion transport model, the diffusion coefficients of lithium ions bound to PEO $D_\mathrm{Li}^\mathrm{PEO}$, the diffusion coefficients of lithium ions bound to anions $D_\mathrm{Li}^\mathrm{anion}$, the total diffusion coefficients of lithium ions $D_\mathrm{Li}^{tot}$, the diffusion coefficients of anions $D_\mathrm{TFSI}$ and $D_\mathrm{DFOB}$ and the transference numbers $t_\mathrm{Li}$ in the investigated systems. The unit of the diffusion coefficients $D$ is $10^{-3} \mathrm{nm^2/ns}$.}
		\label{results}
		\centering
		\begin{tabularx}{\textwidth}{p{2cm} X X X X X X X X X X X}
			\toprule
			salt & $N$ & $p_\mathrm{PEO} \; [\%]$ & $\tau_\mathrm{1,eff} \; [\mathrm{ps}]$ & $\tau_3 \; [\mathrm{ns}]$ & $D_\mathrm{Li}^\mathrm{Peo}$ & $D_\mathrm{Li}^\mathrm{anion}$ & $D_\mathrm{Li}^\mathrm{tot}$ & $D_\mathrm{TFSI}$ & $D_\mathrm{DFOB}$ & $t_\mathrm{Li}$ \\ 
			\midrule
			\multirow{1}{\textwidth}{\centering\textbf{Systems without network defects}} \\
			\midrule
			TFSI & 12 & 99.7 & 202 & 326 & 0.48 & 6.86 & 0.50  & 21.2 & & 0.02 \\
			TFSI & 24 & 99.5 & 105 & 175 & 1.81 & 34.38 & 1.96 & 40.6 & & 0.05 \\
			TFSI & 36 & 99.7 & 91 & 166 & 2.82 & 38.46 & 2.93 & 60.5 & & 0.05 \\
			\hline
			DFOB & 12 & 88.9 & 459 & 81 & 1.64 & 4.23 & 1.93  & & 8.5 & 0.18 \\
			DFOB & 24 & 73.8 & 163 & 50 & 4.65 & 3.86 & 4.46  & & 10.1 & 0.30 \\
			DFOB & 36 & 86.2 & 121 & 44 & 8.01 & 9.86 & 8.06  & & 21.1 & 0.28 \\
			\hline
			TFSI/DFOB & 12 & 97.2 & 291 & 155 & 0.97 & 4.96  & 1.08 & 15.7 & 10.9 & 0.07 \\
			TFSI/DFOB & 24 & 96.0 & 130 & 109 & 3.43 & 13.29 & 3.82 & 31.9 & 19.2 & 0.13 \\
			TFSI/DFOB & 36 & 97.6 & 104 & 92 & 4.68 & 19.01 & 5.02 & 46.9 & 28.0 & 0.12 \\
			\midrule
			\multirow{1}{\textwidth}{\centering\textbf{Systems with network defects}} \\
			\midrule
			TFSI & 24 & 99.7 & 103 & 172 & 2.61 & 35.90 & 2.70 & 48.4 & & 0.05 \\
			
			DFOB & 24 & 75.5 & 156 & 45 & 7.04 & 10.39 & 7.52  & & 17.3 & 0.30 \\
			
			TFSI/DFOB & 24 & 97.1 & 125 & 93 & 4.66 & 17.78 & 5.04 & 42.8 & 25.3 & 0.12 \\
			\midrule
			\multirow{1}{\textwidth}{\centering\textbf{Non-cross-linked systems}} \\
			\midrule
			TFSI/DFOB & 36 & 97.9 & 80  & 43.7 & 28.79 & 45.25 & 29.13 & 124.4 & 66.4 & 0.22 \\
			* & 36 & 97.9 & 80 & 43.7 & 5.97 & 45.25 & 6.78 & 124.4 & 66.4 & 0.06 \\
			
			\bottomrule
			
			\multirow{1}{\textwidth}{*The diffusion coefficients and transference number of lithium ions were calculated using the reduced diffusion coefficient $D_\mathrm{Li}^\mathrm{no\;C.O.M}$ (see Equation \ref{Eq_D_com}) for lithium ions coordinated to PEO chains. }
		\end{tabularx}
	\end{table*}
	\vspace{5cm}
	
	\begin{equation}
		D_\mathrm{Li}^\mathrm{tot} = D_\mathrm{Li}^\mathrm{PEO} \cdot p_\mathrm{PEO} + D_\mathrm{Li}^\mathrm{anion} \cdot (1-p_\mathrm{PEO}).
		\label{D1}
	\end{equation}
	In this equation, $D_\mathrm{Li}^\mathrm{PEO}$ is the diffusion coefficient of lithium ions that are bound to any PEO chain, which includes lithium ions that are bound to a PEO chain and one or multiple anions at the same time and $D_\mathrm{Li}^\mathrm{anion}$ is the diffusion coefficient of lithium ions that are bound to any anion, but not to a PEO chain. $p_\mathrm{PEO}$ is the percentage of lithium ions that are at least bound to one PEO chain. Note that we assume that there are no unbound lithium ions, which is reasonable as could be seen from coordination distributions.
	
	All diffusion coefficients are listed in Table \ref{results}. We observe that $D_\mathrm{Li}^\mathrm{tot}$ is mainly determined by $D_\mathrm{Li}^\mathrm{PEO}$ due to the large percentage of lithium ions that are coordinated trough at least one PEO chain in TFSI systems and similar diffusion coefficients of $D_\mathrm{Li}^\mathrm{PEO}$ and $D_\mathrm{Li}^\mathrm{anion}$ in DFOB systems.	
	Furthermore, $D_\mathrm{Li}^\mathrm{tot}$ increases with increasing chain length of PEO polymers and $D_\mathrm{Li}^\mathrm{tot}$ is larger in pure DFOB salt systems than in pure TFSI salt systems. 
	Relating the obtained $D_\mathrm{Li}^\mathrm{tot}$ to the MSD of the lithium ions at larger time scales shown in Figure \ref{total_msd}, we see that both follow the same trends. The MSD and $D_\mathrm{Li}^\mathrm{tot}$ increase with increasing chain length and both are larger for DFOB systems than for TFSI systems with the dual salt system intermediate.

	The largest diffusion coefficient was found for the pure DFOB salt system with 36 monomers per PEO chain. Comparing all the results, we conclude that increasing the chain length in pure TFSI salt systems is more effective than in pure DFOB salt systems, which is in line with our findings above. (That is, the transport is more PEO based in TFSI systems.) We observe from the data in Table\;\ref{results} that the effect of chain length is most pronounced for short chains, which is comparable to earlier findings on systems with linear PEO chains showing a well-defined limit for infinitely long chains \cite{Chattoraj.2015}. However, contrary to the earlier findings, we are also for short chains in the limit of infinitely long chains when considering the center of mass motion of them, which therefore does not contribute to the lithium ion motion in cross-linked polymers significantly. Furthermore, the diffusion coefficients are decreasing with an increasing chain length in linear polymers, whereas they are increasing in the here presented cross-linked polymers. The significantly decreasing center of mass motion of polymer chains with increasing chain length in linear unconstrained polymers leads to this decrease. In cross-linked polymers only the motion of lithium ions along the polymer backbone and ion jumping are significantly affected, which are both accelerated for an increasing chain length. For both cases the diffusion coefficients seem to plateau.
	
	To compare our results to experiments we calculated the transference number $t_\mathrm{Li}$, which can be determined from experiment. In principle, it can be exactly calculated as
	
	\begin{equation}
		t_\mathrm{Li}=\frac{I_\mathrm{Li}}{I_\mathrm{Li}+I_\mathrm{anions}}=\frac{\sigma_\mathrm{Li}}{\sigma_\mathrm{Li}+\sigma_\mathrm{anions}}.
	\end{equation}
	$I_\mathrm{i}$ is the current and $\sigma_i$ is the conductivity of species i.
	For sufficient long time scales a plot of $t_\mathrm{Li}(t)$ should reach a plateau value which is the transference number $t_\mathrm{Li}$. Unfortunately, our simulations are too short to reach a plateau value due to large uncertainties related to the calculation of the conductivity in MD simulations. Because longer simulations are computationally to expensive, we approximated the transference number as
	
	\begin{equation}
		t_\mathrm{Li}=\frac{D_\mathrm{Li}}{D_\mathrm{Li}+D_\mathrm{anions}}
	\end{equation}
	by neglecting the correlated motion of ions. Instead of the conductivities we only need to calculate the MSD of the anions. Within our simulation time the dynamics of anions becomes diffusive, which means that we are able to calculate the diffusion coefficient $D_\mathrm{anions}$ in a straightforward manner through the Einstein equation. The results are listed in Table \ref{results}.
	We observe for all systems a lower transference number for small PEO chains ($N=12$) than for the longer chains. Interestingly, the transference number for $N=36$ is similar to the transference number for $N=24$. This fits our expectation that there should be a plateau value for infinitely long chains. To prove this, further investigations would be needed. 
	Nevertheless, the effect of the chain length is minor in comparison to the effect of the lithium salt. The transference number is approximately 6 times higher in pure DFOB salt systems than in pure TFSI salt systems, however, ionic correlations which have been neglected here might be more pronounced in the former due to the stronger ion pairing, which would impede the lithium dynamics and thus diminish the true transference number.

	\section{Comparison to systems with network defects and non-cross-linked systems}
	So far, idealized initial network structures were used, in which every PEO chain is connected at each of its ends to a PE chain. Although this seems reasonable based on the findings in Section \ref{Sec:network_structure}, real network structures are likely not perfect. To estimate the differences between an idealized structure and a more realistic structure, we performed simulations on systems in which 25\% of the PEO chains are connected only at one end to a PE chain, as described in Section \ref{sec_MD_simulations}. We analysed these systems as described above and observed slight differences in systems structure and dynamics.
	
	As expected, we observe a broader distribution of the angles between the end-to-end vectors of two neighbouring PEO-chains, which is simply caused by a higher number of possible orientations for PEO chains that are only connected to one PE chain.
	The coordination of lithium ions through ether oxygens or anions is not affected significantly by this, as also observed from the agreement of Figure \ref{coordination} with our previous results \cite{Shaji.2022}. 
	Whereas the structural changes may help to explain the dynamics, we were mainly interested in the dynamics itself. In general, we observe an increased dynamics for systems with network defects, including shorter relaxation times and larger MSDs, caused by a more flexible network structure in network systems with defects. The calculated values are listed in Table \ref{results}.
	Comparing the results to the results of systems without network defects, we note that the time a lithium ion needs to move along a chain ($\tau_\mathrm{1,eff}$) decreases slightly. The one dimensional MSD $\langle \Delta n^2(t) \rangle$ is only slightly increased for systems with network defects in comparison to networks without defects for time scales below 10 ns. For larger time scales we observe a larger increase (see Figure \ref{SI:fig_MSD_linear_def}). For $\tau_3$ we observe only slight differences between systems with and without network defects, meaning that primarily the dynamics of mechanism 1 and 2 are increased in the former.
	We observe larger diffusion coefficients $D$ for all salt species, but they are still on the same order of magnitude as for systems without network defects. 
	Interestingly, we found no significant effect on the transference number $t_\mathrm{Li}$. This means in fact that the dynamics of lithium ions and the dynamics of anions increase in such a way that $t_\mathrm{Li}$ is nearly constant. This finding is very important for the comparison with experimental data, as the dynamics of the system is essentially insensitive to the presence of defects.
	
	Comparing the results of cross-linked-polymers to classical polymer melts (see Table\;\ref{results}) we observe a very drastic decrease of the relaxation times in the polymer melt. This could be explained as described above by more possible orientations and conformations of the PEO chains. The diffusion coefficient for lithium ions is one order of magnitude larger than in  the cross-linked polymers. However, it cannot be directly compared, because in non cross-linked polymers the center of mass motion of chains is very large, having a huge impact on total diffusion coefficient due to mechanism 2, whereas in cross-linked-polymers it is negligibile. If for a first approximation the center of mass motion of PEO chains $D_\mathrm{PEO}^\mathrm{C.O.M}$ is subtracted and the diffusion coefficient
	\begin{equation}
		\label{Eq_D_com}
		D_\mathrm{Li}^\mathrm{no\;C.O.M} = D_\mathrm{Li}^\mathrm{PEO} - D_\mathrm{PEO}^\mathrm{C.O.M}
	\end{equation}
	is calculated, which is comparable to the diffusion coefficients of lithium ions in cross-linked polymers (see Table \ref{results}). Therefore, if long linear polymer chains were employed, we would expect that the differences with respect to polymer networks would be less pronounced.
	
	\section{Conclusions}
	In this article, we performed MD simulations on cross-linked PEO electrolytes with different lithium salts and applied a well studied lithium ion transport model for PEO melts to them.
	As the exact structure of the formed networks is unknown, we employed idealized structures. We varied the length of the PEO chains and added the two different pure lithium salts LiTFSI and LiDFOB as well as their mixture to the polymer network, which was motivated by recent experimental findings \cite{Shaji.2022}. 
	First, we analysed the structure of equilibrated networks. It turned out that the structures of single PEO chains in networks could be described as in PEO melts except that in networks less conformations for PEO chains are possible, which leads to a smaller distribution of end-to-end vectors and partly more elongated chains. Regarding all PEO chains in the network, they retain some local order due to the idealized structure. This order becomes lower when network defects are present. The lithium coordination is essentially unaffected by the network structure. 
	Next, we analysed the transport of lithium ions according to the lithium ion transport model with respect to the PEO chain length and employed lithium salt. 
	In general we observed increased dynamics in systems with longer PEO chains, in which chain end effects become less relevant. However, the total lithium ion dynamics is affected mainly by the chosen lithium salt. 
	This could be explained by an additional transport mechanism for lithium ions, which emerges when using LiDFOB. DFOB is able to decouple the lithium ion from the polymer chain. Movement of lithium ions together with anions is fast in comparison to movement of lithium ions together with the slow polymers, leading to increased dynamics.
	Additionally, the decoupling mechanism supports jumps of lithium ions from one PEO chain to another PEO chain.
	The ability of DFOB to decouple lithium ions is based on the one hand on a larger average coordination time of DFOB to lithium ions in contrast to TFSI. On the other hand DFOB is able to form large and stable clusters with lithium ions.
	Often two or more DFOB molecules coordinate to one lithium ion at the same time, enabling them to decouple $\mathrm{Li^+}$ cooperatively. However, the formation of very large clusters may slow down the ion transport significantly, since the ions would become trapped inside of it. In comparison, in TFSI systems often only one TFSI molecule coordinates to a given lithium ion. Systems in which both salts are employed show intermediate cluster sizes, in which lithium ions are not trapped, but can still be decoupled from PEO chains efficiently by DFOB to enable higher transfer rates of them between different PEO chains. 
	Furthermore, non-trivial effects for the cation-anion crosscorrelation terms might play a role when calculating the ionic conductivity, which can explain the improvement of the dual salt system compared to single salt systems. We propose that further research in this field can improve the understanding of multi salt systems significantly.
	
	Finally, we calculated the transference number $t_\mathrm{Li}$, which is much higher in pure LiDFOB systems than in pure LiTFSI systems. Interestingly, for long PEO chains ($N = 24$ and $N=36$), there is only a small impact of the chain length on $t_\mathrm{Li}$. Moreover, even network defects only have a marginal impact (within the statistical accuracy) on $t_\mathrm{Li}$.
	This emphasizes that even the comparatively simple idealized structures in our MD simulations nonetheless yields reasonable results, which should be comparable to experiments. This may be helpful for future research of cross-linked polymer electrolytes, regarding theoretical, but also experimental considerations in order to understand transport mechanisms.
		
	\printbibliography
	\clearpage

\end{document}


\maketitle

\clearpage

	\section{Simulated Systems}
	\begin{table}[h]
	\centering
	\caption{Composition of all simulated systems. The chain length as well as the number of PEO chains $N_\mathrm{PEO}$, the number of LiTFSI ion pairs $N_\mathrm{TFSI}$ and the number of LiDFOB ion pairs $N_\mathrm{DFOB}$ are given. Furthermore, the sizes of the simulation box in x and y direction $b_x$ and $b_y$ and the size in z direction $b_z$ are given. The simulation time of the production run $t_\mathrm{sim}$ is also tabulated. The column \textit{cross-linked} gives information whether the PEO chains are cross-linked (\textit{yes}) or not (\textit{no}) and if network defects are present (\textit{defect}). The (r) indicates that the systems are used as reference systems.}
	\begin{tabularx}{\textwidth}{p{0.8cm} p{2.1cm} X X X X X X X X}
		\toprule
		Index & salt & cross-linked & $N$ & $N_\mathrm{PEO}$ & $N_\mathrm{TFSI}$ & $N_\mathrm{DFOB}$ & $b_x$ / $b_y$ [nm] & $b_z$ [nm] & $t_\mathrm{sim}$ [ns]\\
		\midrule
		 1 & TFSI      & yes     & 12 &  64 &  54 &   0 & 3.58 &  5.96 &  550 \\
		 2 & TFSI      & yes     & 24 &  48 &  82 &   0 & 4.95 &  4.54 & 1000 \\
		 3 & TFSI      & yes     & 36 &  64 & 163 &   0 & 6.10 &  5.90 &  860 \\
		 \midrule
		 4 & DFOB      & yes     & 12 &  64 &   0 &  54 & 3.51 &  5.61 & 1000 \\
		 5 & DFOB      & yes     & 24 &  48 &   0 &  82 & 5.02 &  3.98 & 1000 \\
		 6 & DFOB      & yes     & 36 &  64 &  82 & 163 & 5.90 &  5.69 & 1000 \\
		 \midrule
		 8 & TFSI/DFOB & yes     & 12 &  64 &  32 &  22 & 3.59 &  5.69 & 1000 \\
		 9 & TFSI/DFOB & yes     & 24 &  48 &  48 &  34 & 4.99 &  4.99 & 1000 \\
		10 & TFSI/DFOB & yes     & 36 &  64 &  96 &  67 & 6.08 &  5.71 &  779 \\
		\midrule
		11 & TFSI      & defect  & 24 &  64 & 110 &   0 & 4.98 &  5.94 & 1000 \\
		12 & DFOB      & defect  & 24 &  64 &   0 & 110 & 4.85 &  5.76 & 1000 \\
		13 & TFSI/DFOB & defect  & 24 &  64 &  65 &  45 & 4.90 &  5.97 & 1000 \\
		\midrule
		14 & TFSI/DFOB & no      & 36 &  40 &  60 &  42 & 5.08 &  5.08 & 1000 \\
		\midrule
		15 & TFSI      & yes (r) & 24 & 160 & 272 &   0 & 5.00 & 14.64 &  400 \\
		16 & DFOB      & yes (r) & 24 & 160 &   0 & 272 & 4.95 & 13.67 &  400 \\
		17 & TFSI/DFOB & yes (r) & 24 & 160 & 160 & 112 & 4.99 & 14.17 &  400 \\
		\bottomrule

	\end{tabularx}
	\end{table}
	
	\clearpage

	\section{Comparison of Different System Sizes}
	\begin{figure}[h]
		\centering
		\includegraphics[width=0.7\linewidth]{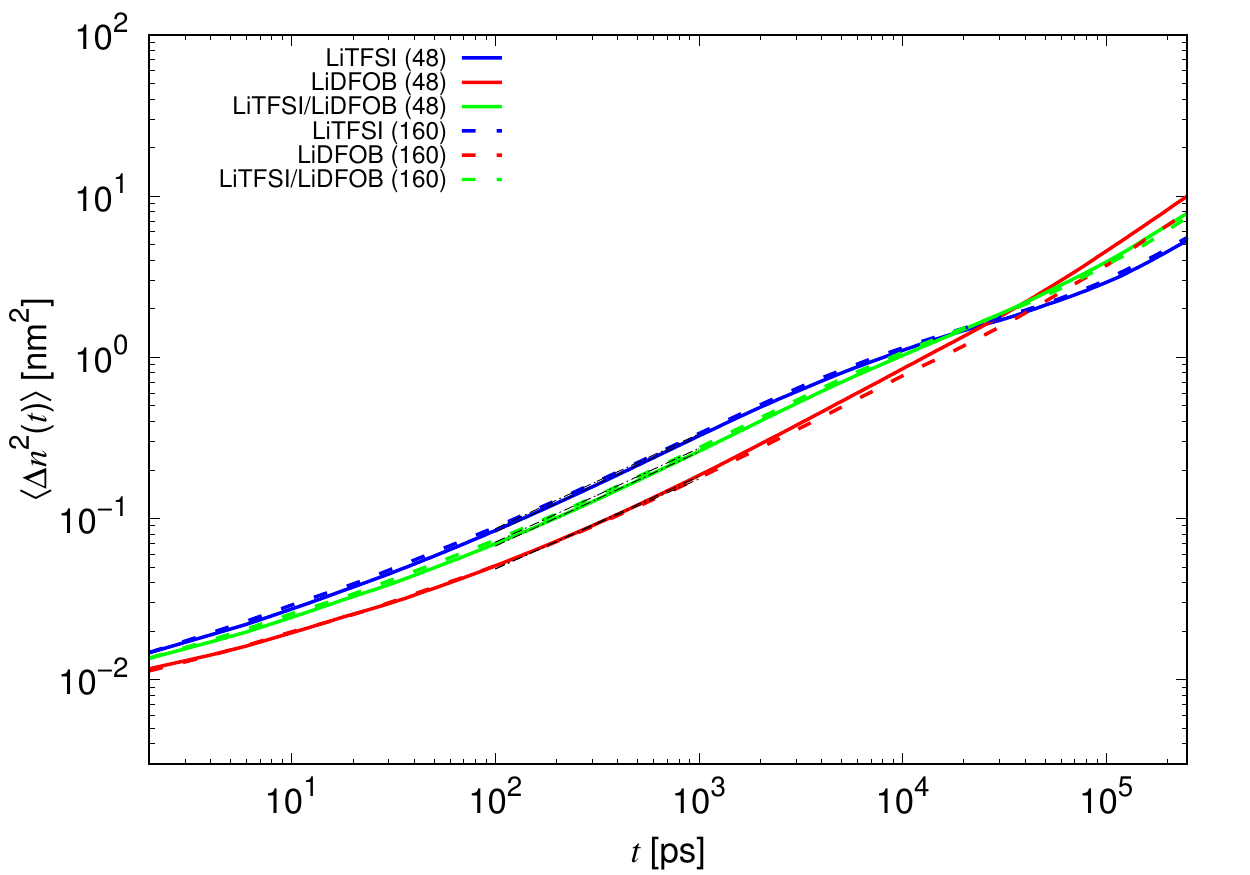}
		\caption{Effect of the simulation box size on the lithium ion dynamics. The MSDs of lithium ions in the analyzed systems are compared to the MSDs of the lithium ions in larger reference systems. As can be seen from the plot, the MSDs for pure LiTFSI salt system and the mixed-salt system are in good agreement for different total number of PEO chains in the system. The MSD in the pure LiDFOB salt system is slightly lower for the larger reference system at time scales larger than 1 ns, due to formation of large clusters, which decrease the dynamics.}
		\label{Si:fig_compare}
	\end{figure}
	
	\clearpage

	\section{Verification of Equilibrium state}
	\begin{figure}[h]
	\centering
	\includegraphics[width=0.7\linewidth]{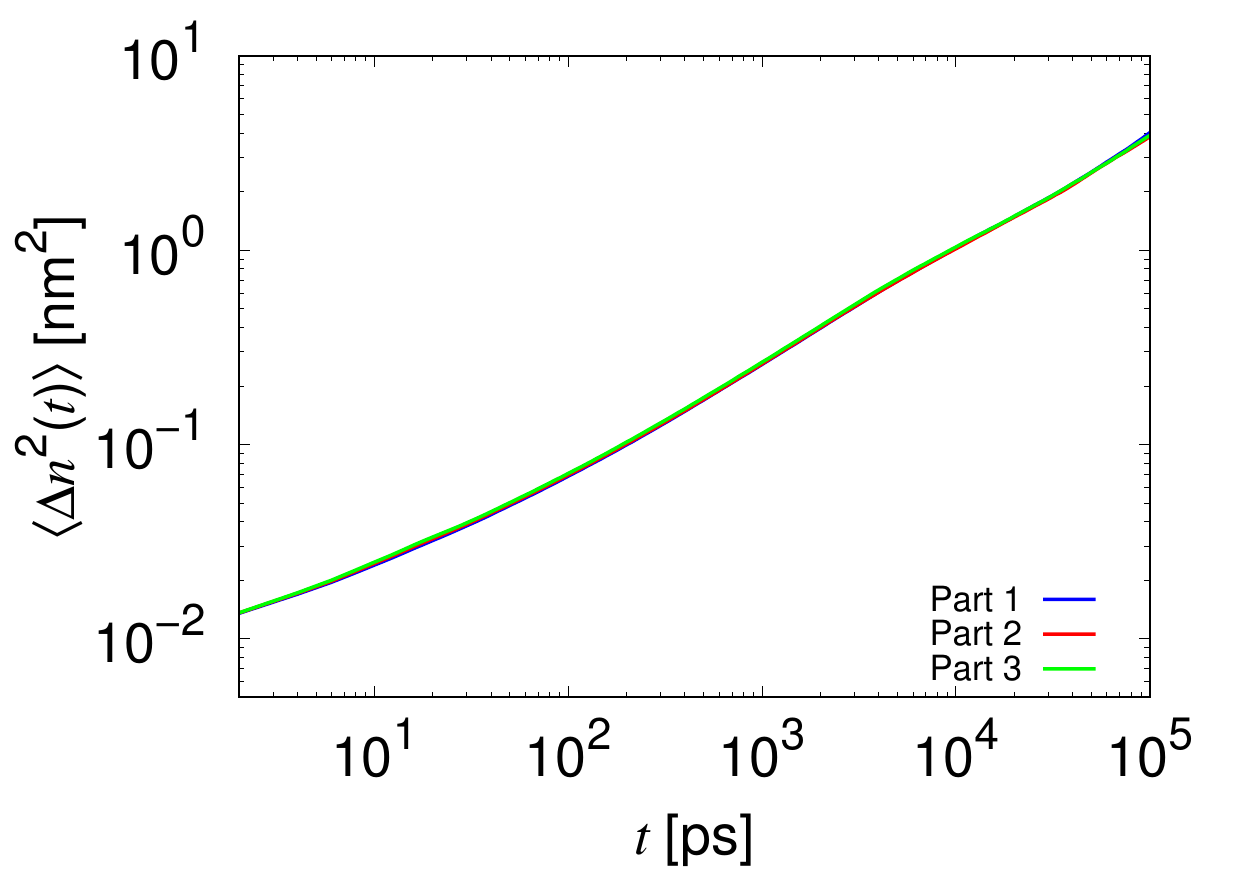}
	\caption{Exemplary, the MSD of the lithium ions is shown for a dual salt system with 24 PEO monomers per chain. The trajectory is splitted in 3 parts and for each part the MSD is calculated separately to make sure the system is equilibrated. The good agreement of the MSDs for all three parts is a good indicator that the system is equilibrated. For other systems the plot looks similar.}
	\label{SI:fig_msd_compare}
	\end{figure}
	
	\begin{figure}[h]
	\centering
	\includegraphics[width=0.7\linewidth]{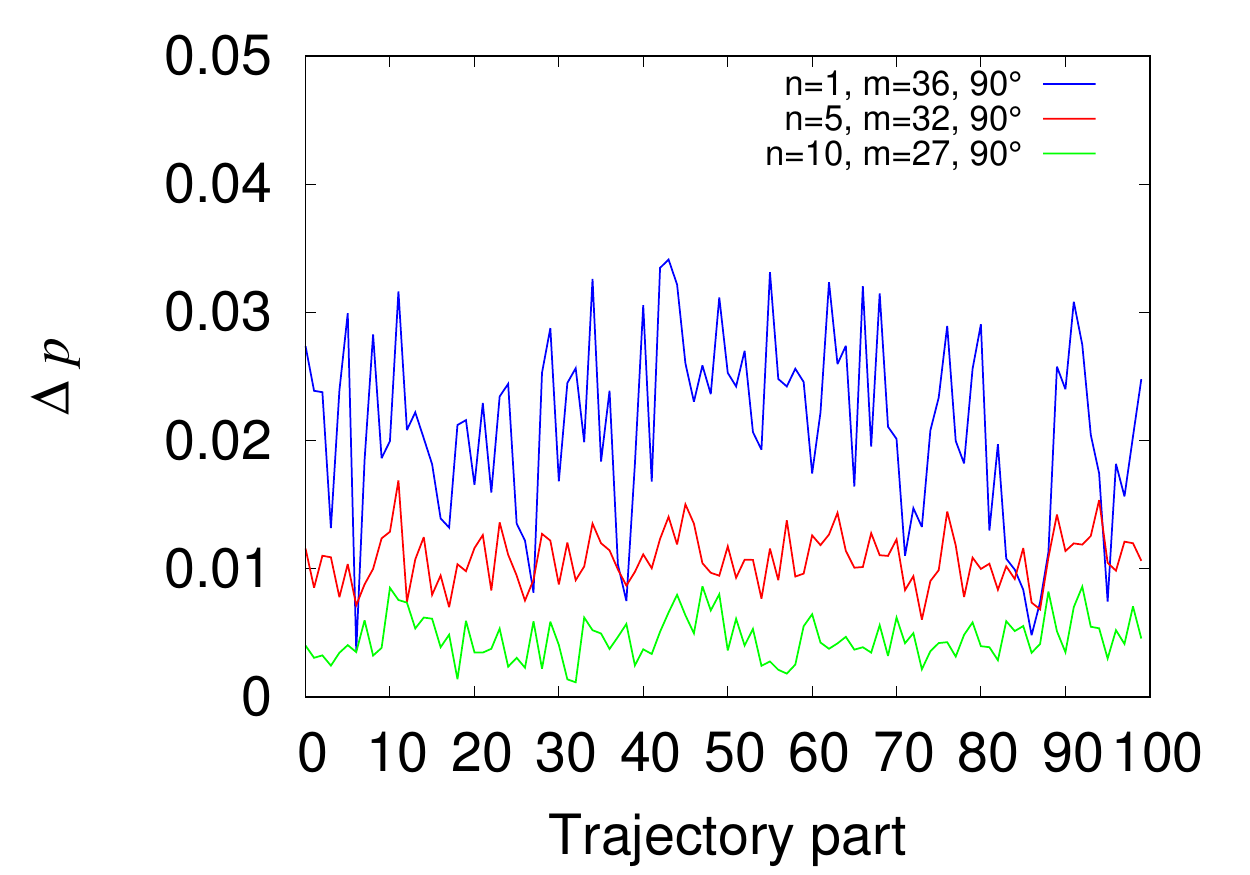}
	\caption{The value of the difference in the probabilities that a polymer chain has a certain orientation to the z-axis and a certain orientation to the x/y-axis $\Delta p = p_\mathrm{z}(\Theta)-p_\mathrm{x,y}(\Theta)$ is shown for an orientation angle of $\Theta=90$° (see also Fig. \ref{SI:fig_network_structure}) in dependency on time for a dual-salt system with a chain length of 36 monomers per chain. The trajectory is splitted in 100 parts, each 10 ns long. Large fluctuations in the relative orientation can be observed especially when following the entire chain ($n=1, m=36$). However, there is no significant change of the investigated parameter over the time, indicating that the system is equilibrated.}
	\label{SI:fig_angle_diff}
	\end{figure}

\clearpage

	\section{Network structure}
	\label{SI:sec_network_structure}
	\begin{figure}[h]
		\centering
		\includegraphics[width=\linewidth]{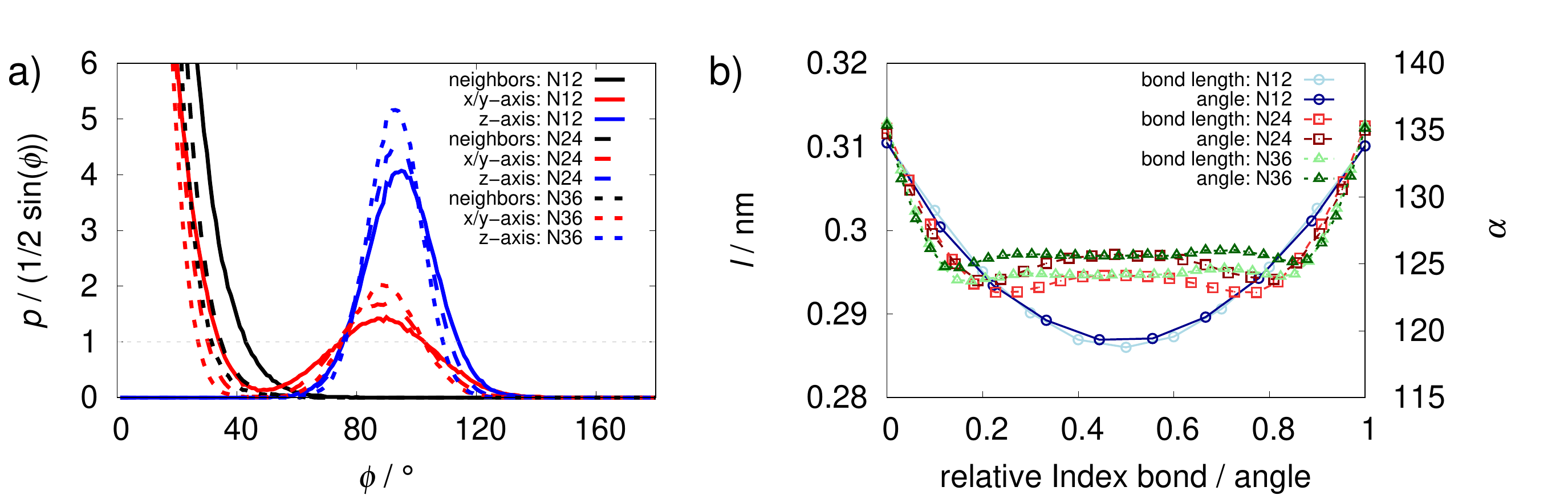}
		\caption{a) The probability distributions of the angles between the PEO chains and unit vectors of the individual spatial dimensions and the angles between a vector in a PEO chain and a vector in a neighbouring PEO chain, which are defined as PEO chains that are adjecent within the initial network structure, divided by $\frac{1}{2}\sin(\phi)$ are shown for systems with different PEO chain length. The gray dashed line illustrates the distribution of orientations in an ideal PEO melt. b) The angles and length between individual monomers of the PEO chains are shown as a function of the relative monomer position within the chain.}
		\label{SI:fig_network_structure}
	\end{figure}

\begin{figure}[h]
		\centering
		\includegraphics[width=0.7\linewidth]{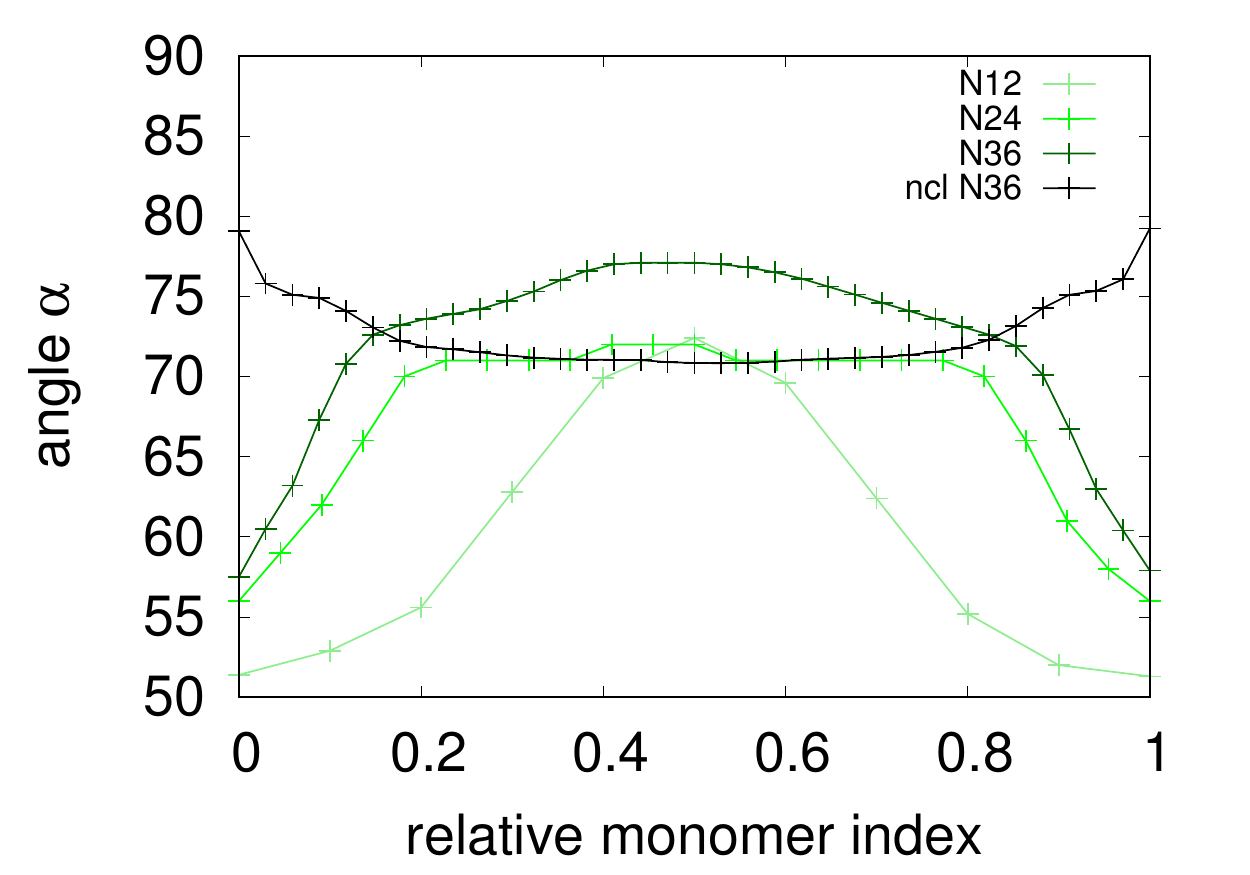}
		\caption{Average angle between the monomer vectors and the end-to-end vector of a PEO chain, dependent on the monomer position within the chain, in the dual salt systems. To compare different chain length, the position is given as an relative monomer index between 0 and 1.}
		\label{SI:fig_angles_monomer}
	\end{figure}

	The analysis, shown in Figure \ref{SI:fig_network_structure}b, reveals that the average angle $\alpha$ between two monomer vectors, connecting the oxygen atoms of adjecent monomers, near the chain end \mbox{($\alpha \approx 128^{\circ}-135^{\circ}$)} is higher than in the middle part \mbox{($\alpha \approx 126 ^{\circ}$)} indicating that the chains tend to be more coiled in the center. For short chains $\alpha$ is even lower in the middle of the chain. This is also supported by the distribution of the angles between the individual monomer vectors and the end-to-end vector of the whole PEO chain (Figure \ref{SI:fig_angles_monomer}).
	We observe that the angle in the network structures is lower at the chain end and higher at the center. With increasing chain length the angle becomes almost perpendicular. Interestingly, in non-cross-linked PEO chains the angle is larger at the chain's end than at the center. 
	
\clearpage
	
	\section{End-to-End-Length}
	\begin{figure}[h]
	\centering
	\includegraphics[width=0.7\linewidth]{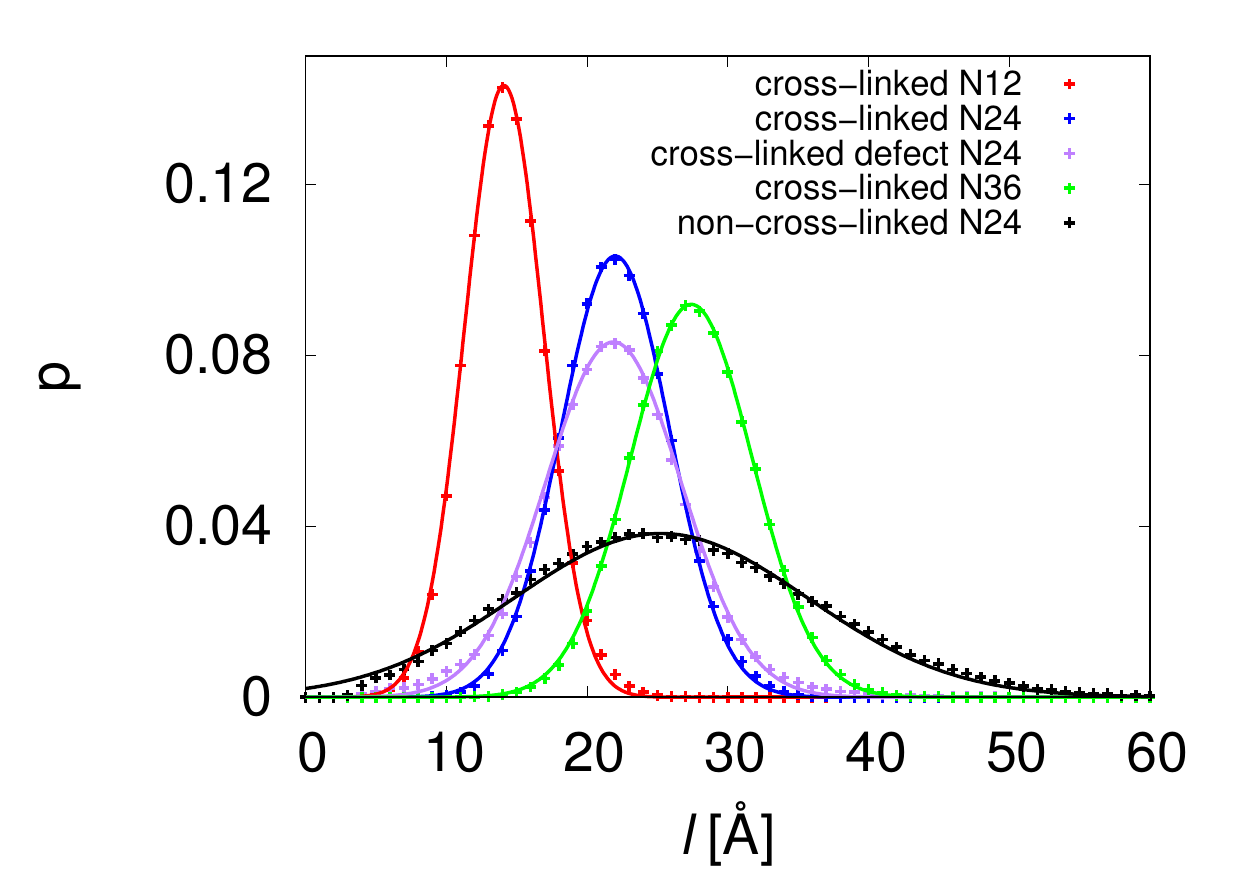}
	\caption{Distribution of the length of the end-to-end vectors of PEO chains in fully cross-linked, in partially cross-linked and in non-cross-linked PEO polymers. The straight line is a Gaussian fit.}
	\label{SI:fig_length}
	\end{figure}
	
	\begin{table}[h]
	\centering
	\caption{From a Gaussian fit obtained parameters for the End-to-end chain length $R_\mathrm{end}$ of the PEO polymers and for the standard deviation $\sigma$.}
	\begin{tabularx}{\textwidth}{p{5cm} X X}
	\toprule
	System & $R_\mathrm{end}$ [nm] & $\sigma$ [$\angstrom$] \\
	\midrule
	N12 cross-linked 			& 14.1 & 2.8 \\
	N24 cross-linked 			& 22.0 & 3.9  \\
	N24 partially cross-linked 	& 21.9 & 4.8 \\
	N36 cross-linked 			& 27.4 & 4.3 \\
	N36 non-cross-linked 		& 25.3 & 10.4 \\
	\bottomrule
	
	\end{tabularx}
	\end{table}

\clearpage	

	\section{Radial distribution functions}
	\begin{figure}[h]
		\centering
		\includegraphics[width=\linewidth]{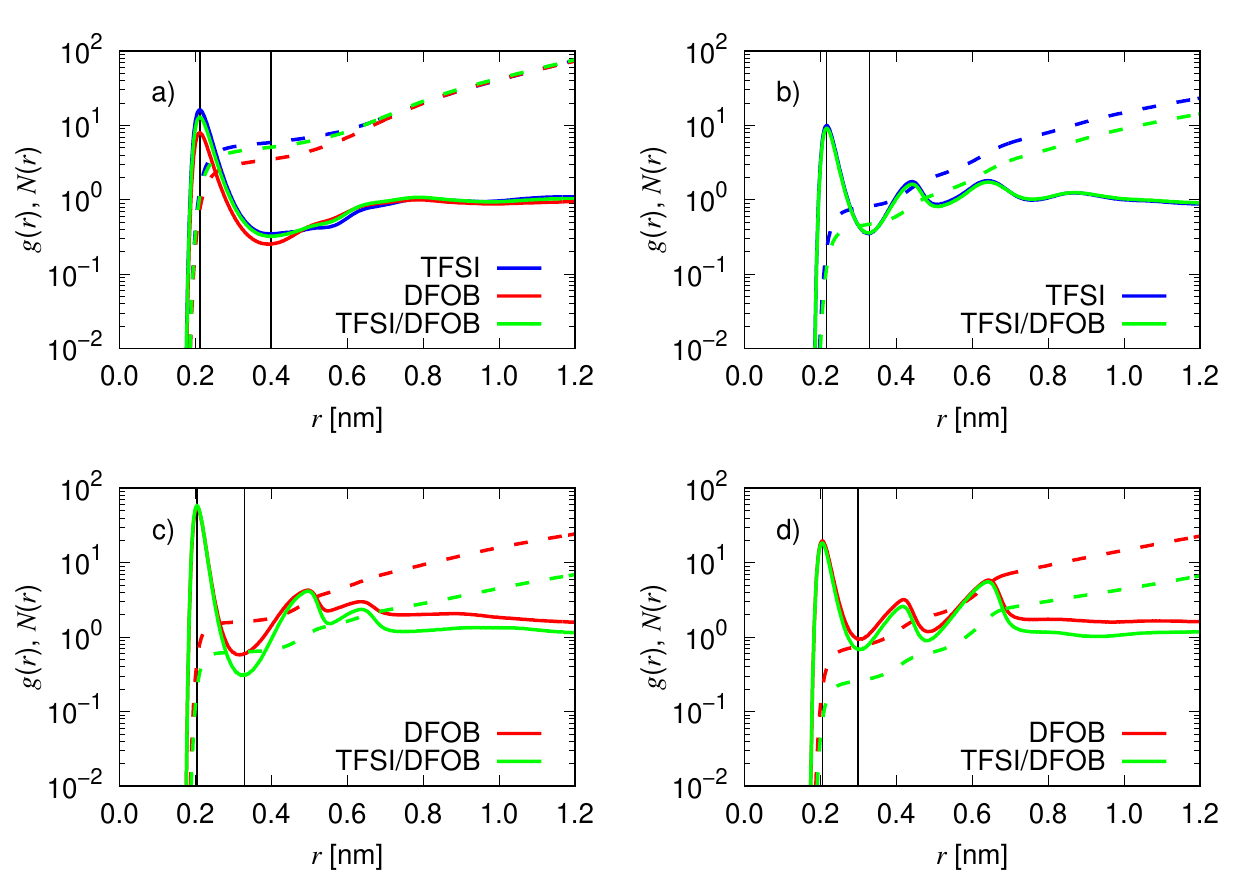}
		\caption{Radial distribution functions $g(r)$ for a) Li and PEO ether oxygen, b) Li and TFSI oxygen, c) Li and DFOB ketone oxygen and d) Li and DFOB fluorine in systems with different lithium salts. The RDFs maximum and the first minimum are flagged by black vertical lines. The dashed lines show the coordination number $N(r)$ at a given distance of the two atoms. }
		\label{SI:fig_rdf}
	\end{figure}
	
\clearpage

	\section{Coordination numbers}
		\begin{figure}[h]
		\centering
		\includegraphics[width=\linewidth]{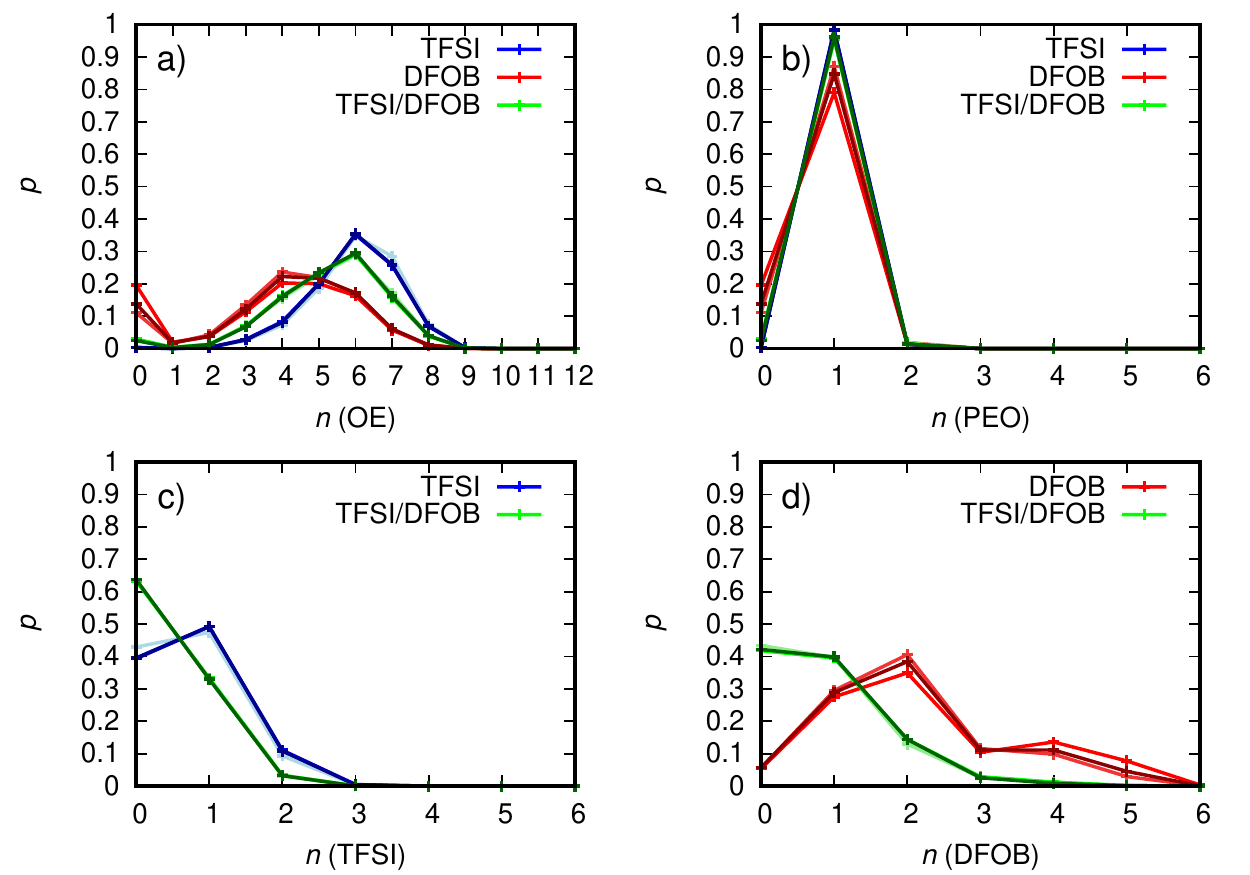}
		\caption{Distribution of coordination numbers in all systems with an idealized network structure. Systems with $N=12$ are colored in light color, systems with $N=24$ are colored as shown in the legend and systems with $N=36$ are colored in dark color.}
			\label{SI:fig_coordination}
	\end{figure}

\clearpage
	
	\section{Cluster}
	\begin{figure}[h]
		\centering
		\includegraphics[width=0.7\linewidth]{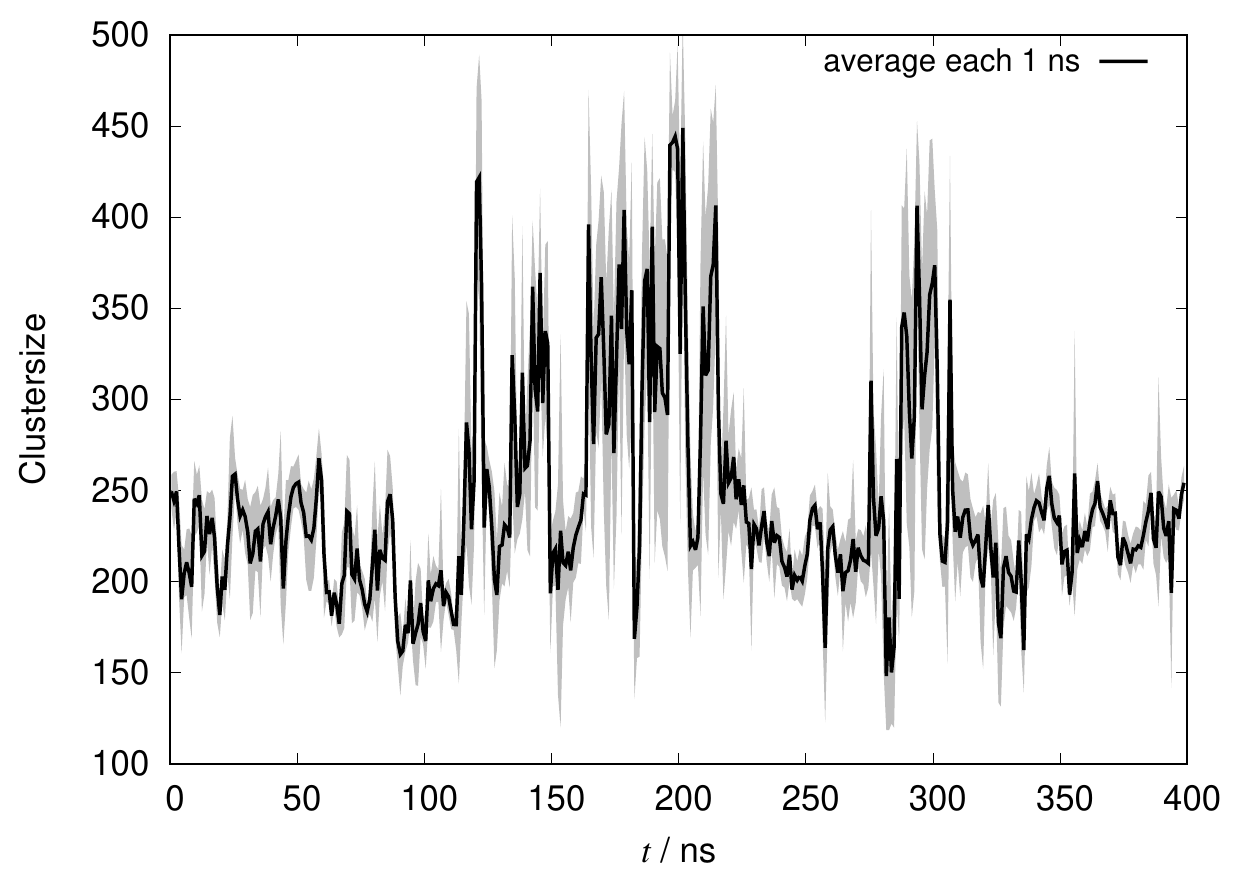}
		\caption{The averaged size of the largest cluster $N_\mathrm{Cluster}$ in the pure LiDFOB system as a function of time $t$ is shown in black. The corresponding standard deviation $\sigma$ is shown in gray.}
		\label{SI:fig_cluster_size}
	\end{figure}
	
	In the following, the procedure to characterize the largest cluster found in the DFOB system is described. First the largest cluster which was formed during the simulation was search over the whole simulation time. Than the size of this cluster was calculated during the simulation time. To quantify the change of the cluster size in a certain time interval we split the trajectory in intervals of 1\;ns each. Then we calculated the average size of the largest cluster $N_\mathrm{Cluster}$ in this interval 
	\begin{equation}
		\langle N_\mathrm{Cluster} \rangle = \frac{1}{N_\mathrm{frames}}\sum_{i=0}^{N_\mathrm{frames}} N_\mathrm{Cluster,i}
	\end{equation}
	where $N_\mathrm{frames}$ is the number of frames in the interval and $N_\mathrm{Cluster,i}$ is the size of the largest cluster at the time frame i in the interval.
	The coefficient of variation $c_v$ was then calculated as
	\begin{equation}
	c_v = \frac{\sigma}{\langle N_\mathrm{Cluster} \rangle}
	\end{equation}
	with $\sigma$ as the standard deviation of the average cluster size. 

	\begin{figure}[h]
		\centering
		\includegraphics[width=0.7\linewidth]{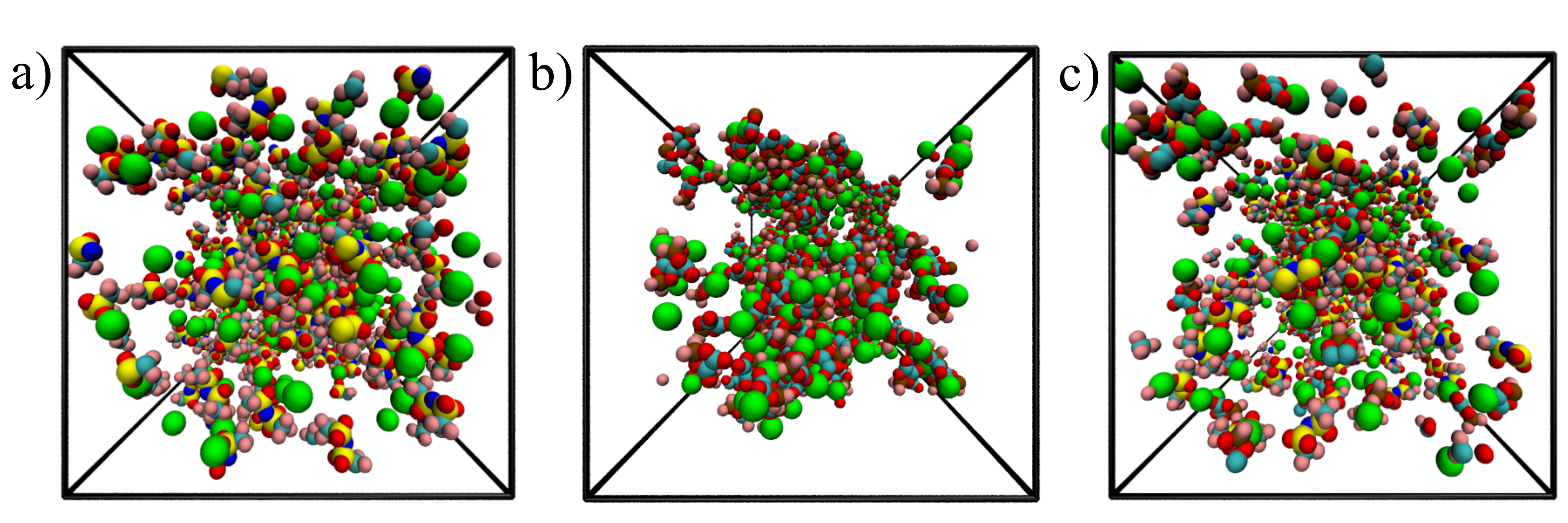}
		\caption{Snapshots from the trajectories of the reference systems. Only anions and lithium ions are shown to demonstrate clustering. Lithium ions are colored green, the atoms of the anions are colored as following: Oxygen (red), Carbon (cyan), Nitrogen (blue), Boron (brown), Sulfur (yellow), Fluorine (pink) a) pure TFSI salt system: only small clusters, b) pure DFOB salt system: a few large clusters, only some smaller clusters, c) mixture salt system: small and medium clusters.}
		\label{SI:fig_cluster_snapshot}	
	\end{figure}

\clearpage
	
	\section{Effective number of inaccessible monomers $a$}
	\label{SI:sec_a}
	\begin{figure}[h]
		\centering
		\includegraphics[width=\linewidth]{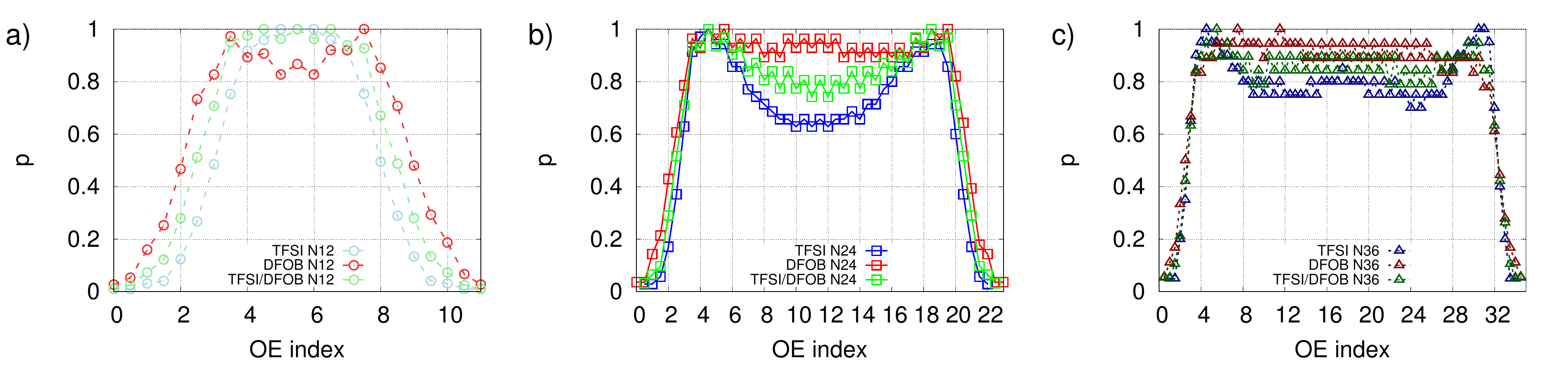}
		\caption{Relative probabilities (in relation to the maximum probability) for lithium ions coordinated to PEO chains to be found at a certain position at the PEO chain. The systems with a) 12, b) 24 and c) 36 monomers per PEO chain are compared with respect to their salt composition.}
		\label{SI:fig_a}
	\end{figure}

	The parameter $a$ was calculated as following. First, the distribution of the lithium ion position at PEO chains as shown in Figure \ref{SI:fig_a} was calculated. Second we defined a relative percentage of 0.7 as a threshold value, which is reasonable, because in most systems in the middle of a chain the relative probability is higher than this value. Only in the pure LiTFSI system with $N=24$ the value is below this threshold value. Next, the relative position $v_\mathrm{0.7,rel}$ at the PEO chain was determined at which the threshold value is reached for the first time, starting from the end of the chain (that has the relative OE index 0). Because of the symmetry of the distribution the obtained value was doubled and rounded to an integer.
	This is the parameter $a$, which defines the region of a chain that can coordinate lithium ions only rarely and therefore reduces the effective chain length that is accessible for lithium ions. 
	Note that we observe no different values for chains with $N=36$ for different salt compositions via this method. Because we observe slight differences for lower indices, nevertheless, we chose the parameter $a$ obtained for the shorter chains.
	Note also that $a$ is approximately the mean coordination number for Li-OE plus 1.
	
	\begin{table}[h]
		\centering
		\begin{tabular}{l r r}
			\toprule
			salt & $N$ & $a$ \\
			\midrule
			TFSI & 12  & 7 \\
			TFSI & 24  & 7 \\
			TFSI & 36  & 6 \\
			DFOB & 12  & 5 \\
			DFOB & 24  & 5 \\
			DFOB & 36  & 6 \\
			TFSI/DFOB & 12 & 6 \\
			TFSI/DFOB & 24 & 6 \\
			TFSI/DFOB & 36 & 6 \\
			\bottomrule
		\end{tabular}
	\caption{Observed values for parameter $a$ in different systems.}
	\label{SI:tab_a}
	\end{table}

\clearpage
	
	\section{Jumping of lithium ions}
	\label{SI:sec_jumping}
	
	\begin{figure}[h]
		\centering
		\includegraphics[width=0.7\linewidth]{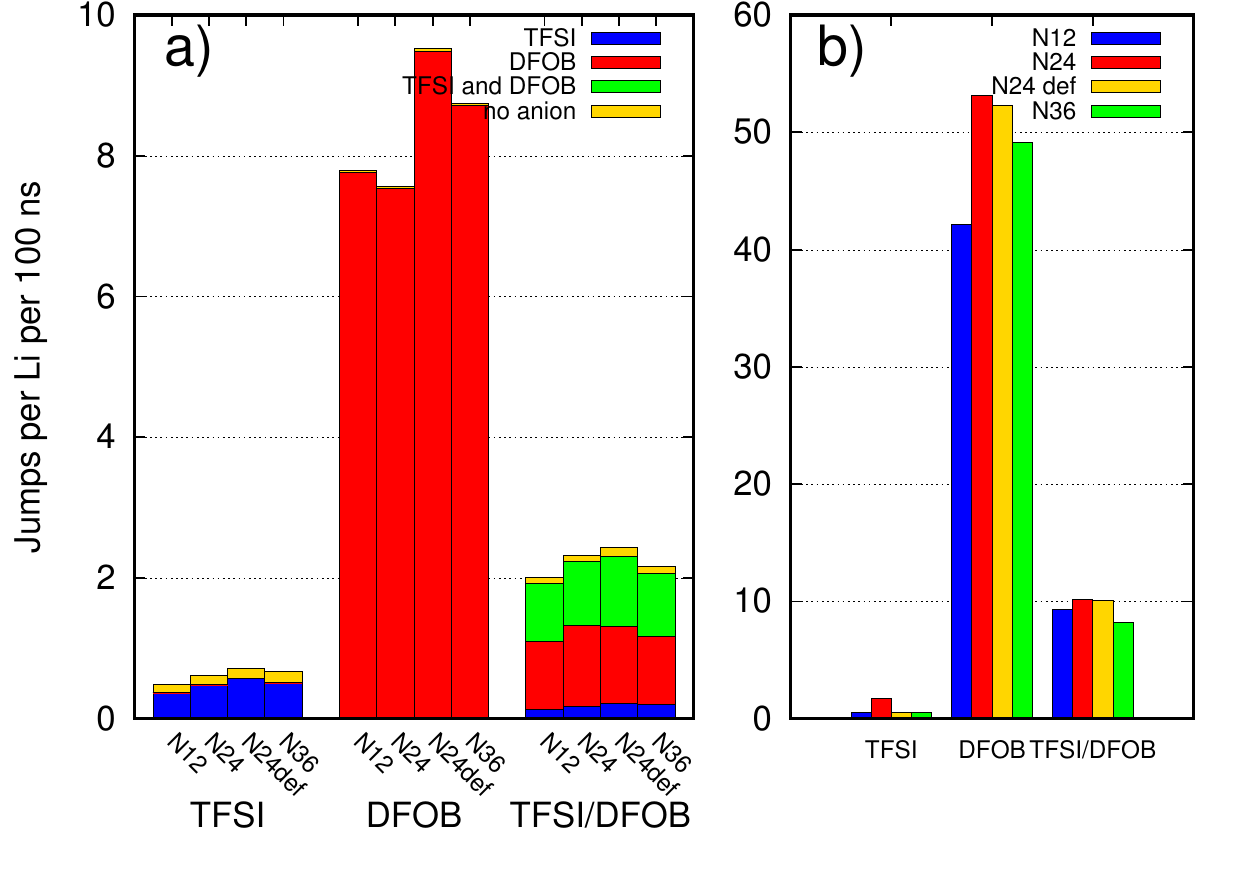}
		\caption{a) Jump rates of lithium ions from one PEO chain to another PEO chain in different systems. The color shows for how many jumps anions are involved and which anions are involved. b) Jump rates of lithium ions from one PEO chain to an anion or vice versa.}
		\label{SI:fig_jumping}
	\end{figure}
	
	We counted the jumps of lithium ions in different systems from one PEO chain to another PEO chain or from a PEO chain to an anion or vice versa. 
	Regarding to the lithium ion transport model it has to be considered that not all jumps are acting as renewal events as in a Random Walk model. Only jumps after which the lithium ion does not jump back should be considered. Usually, we find that the ion jumps back after a few picoseconds. For our analysis we chose a time $\tau = 20\;\mathrm{ps}$. If a lithium ion jumps from one PEO chain to another PEO chain or from a PEO chain to an anion or vice versa and then jumps back within 20~ps the jump is ignored. 
	Furthermore, we define that an anion is involved in the jumping process, if it coordinates to the jumping ion at a certain time interval $[t_\mathrm{jump}-\tau,t_\mathrm{jump}+\tau]$ around the jumping time $t_\mathrm{jump}$, similar to our previous analysis \cite{Shaji.2022}. For this work, $\tau = 20 \mathrm{\;ps}$ was chosen. 	
	
	\begin{figure}[h]
		\centering
		\includegraphics[width=0.7\linewidth]{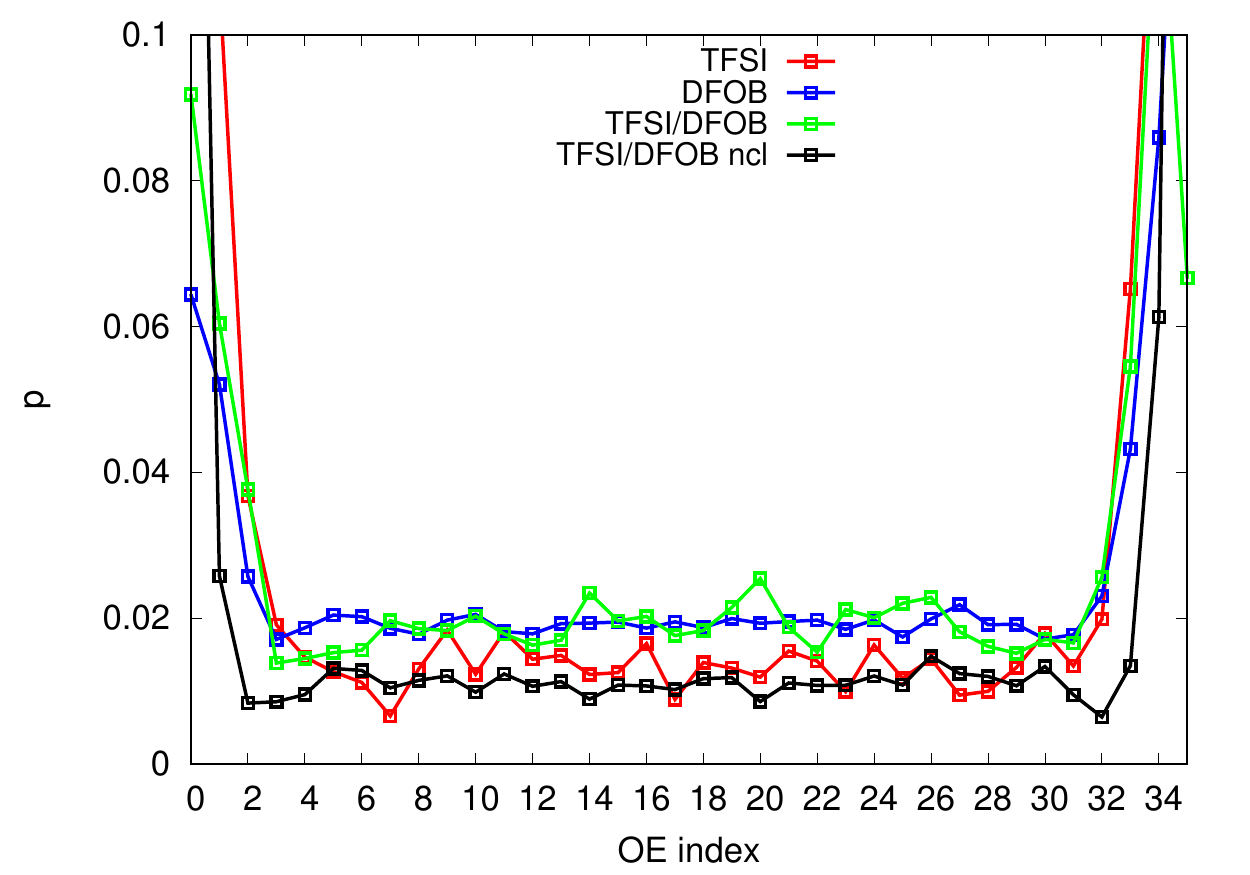}
		\caption{According to Equation \ref{SI:eq_jumping_position} normed probability that a lithium ion jumps from a certain position at a PEO chain (OE index) to another PEO chain. The distribution for jumps to anions looks similar.}
		\label{SI:fig_jumping_position}
	\end{figure}

	Additionally to the jumping rate, we calculated from which position at the PEO chain a lithium ion jumps preferably. Therefore we counted the number of jumps at each ether oxygen index, which was already introduced above in Section \ref{SI:sec_a}. Afterwards we divided the number of jumps by the number of timeframes a lithium ion occupied a certain position to take into account that lithium ions are more often located in the center of the chains. The resulting distribution was then normed, so that the integral is one. The following Equation describes the procedure mathematically
	\begin{equation}
	p_\mathrm{i} = \frac{J_\mathrm{i}}{c_{i}} \text{\huge{/}} \sum_{j}^\mathrm{pos
}  \frac{J_\mathrm{j}}{c_{j}}
	\label{SI:eq_jumping_position}
	\end{equation}
	where $p_i$ is the probability that if a lithium ion jumps it jumps from a certain position (OE index) i. $J_\mathrm{i}$ is the number of jumps from position i and $c_\mathrm{i}$ is the number of timeframes in which a lithium ion was located at position i. pos should indicate the total number of possible positions i.
	
\clearpage

	\section{MSD in x,y,z-dimensions}
	\begin{figure}[h]
		\centering
		\includegraphics[width=0.45\linewidth]{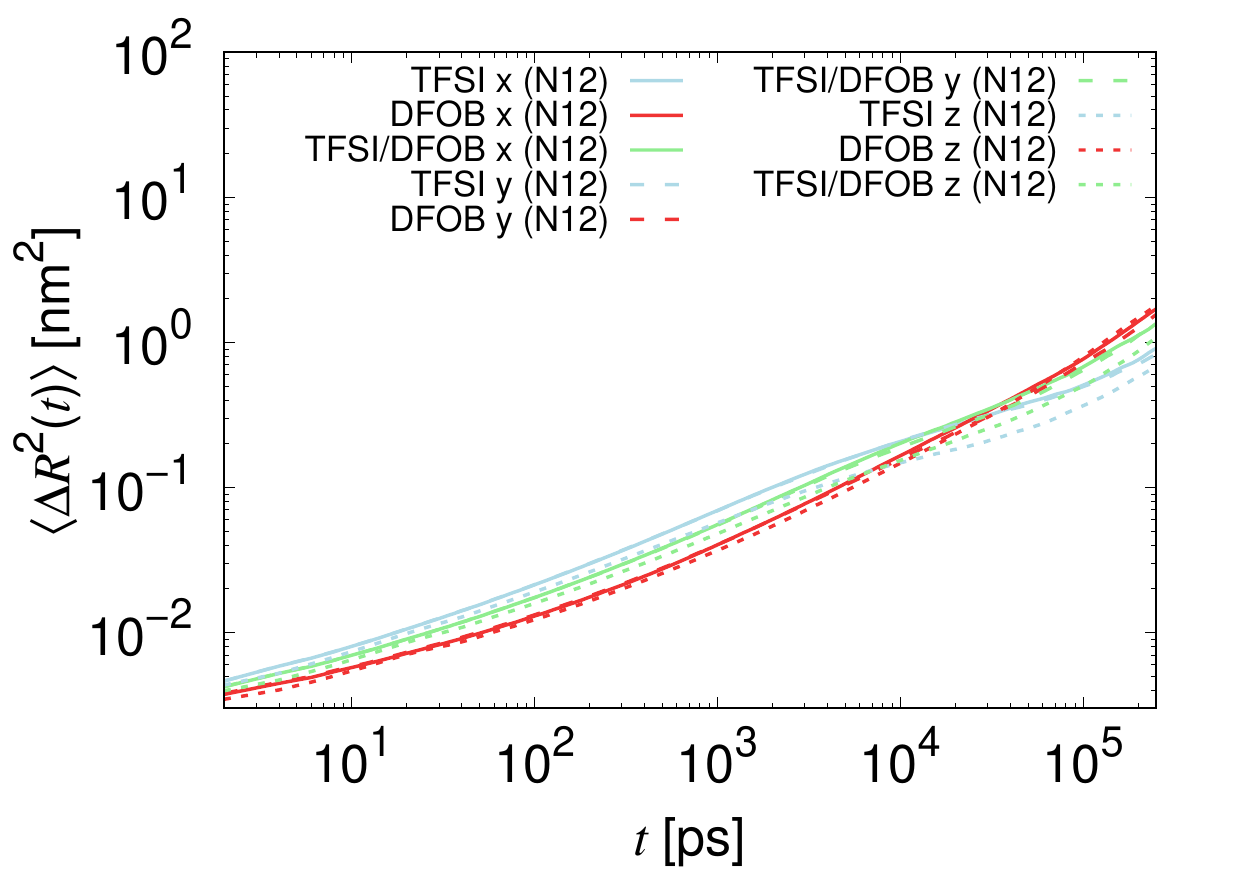}
		\includegraphics[width=0.45\linewidth]{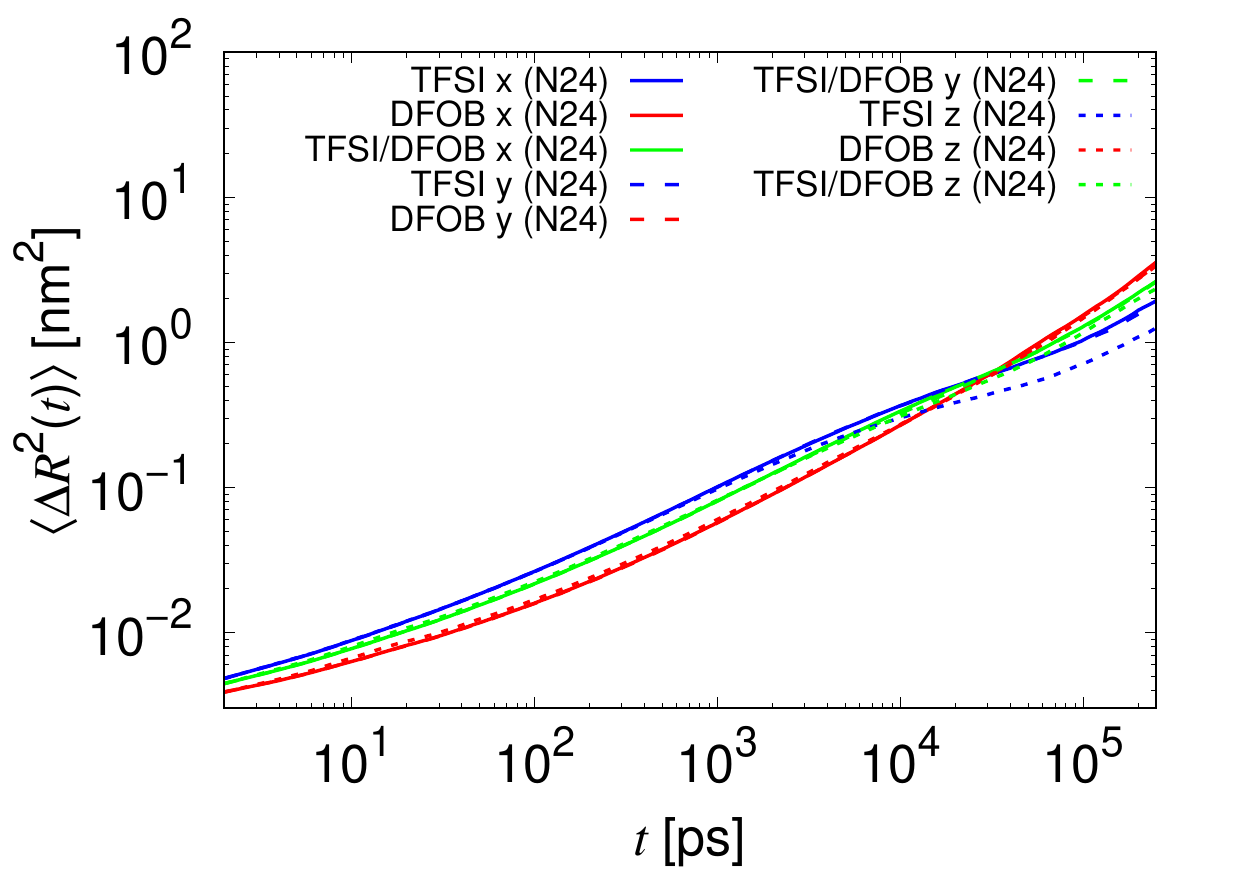}
		\includegraphics[width=0.45\linewidth]{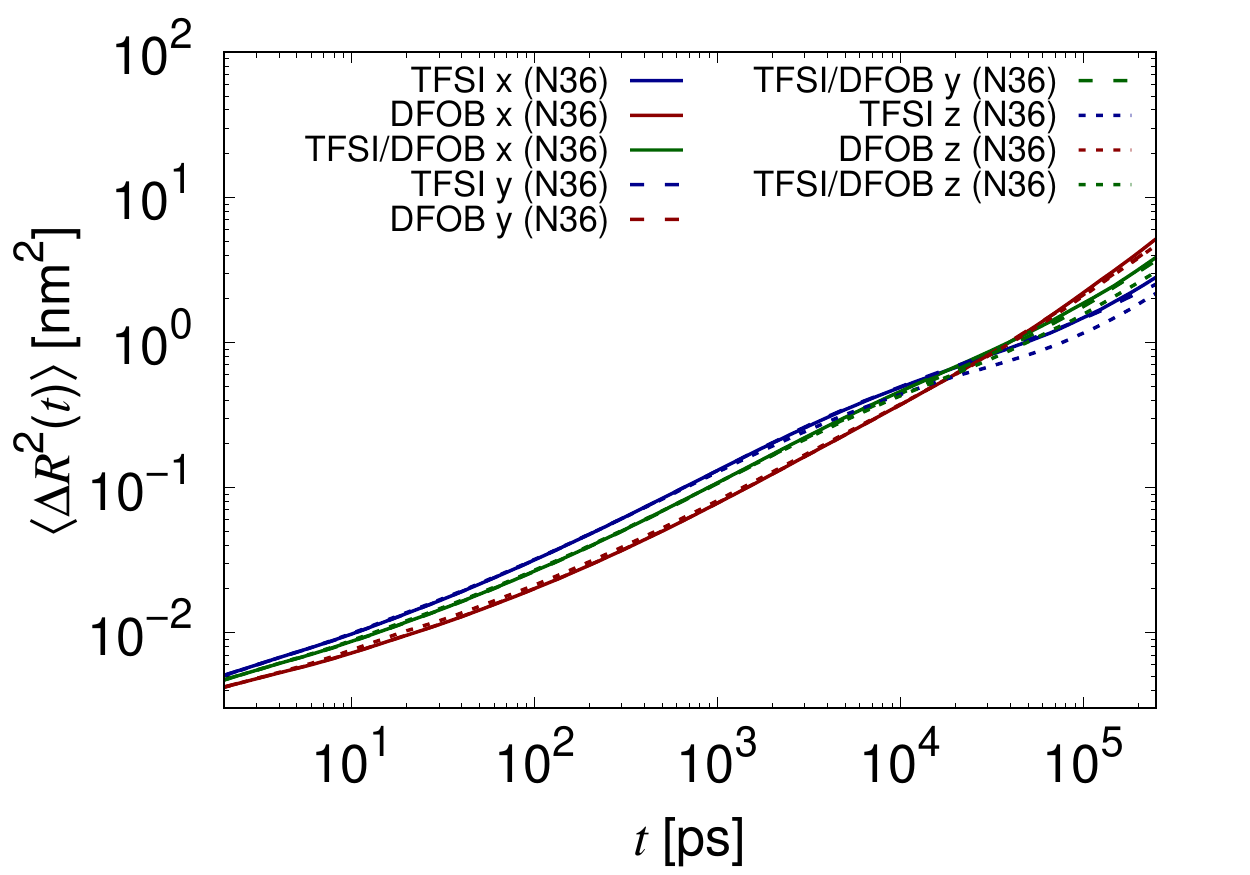}
		\caption{MSDs of lithium ions for different systems according to the spatial directions.}
		\label{SI:fig_msd_direction}
	\end{figure}

\clearpage

	\section{MSD of lithium ions coordinated exclusively to anions}
	
	\begin{figure}[h]
		\centering
		\includegraphics[width=0.7\linewidth]{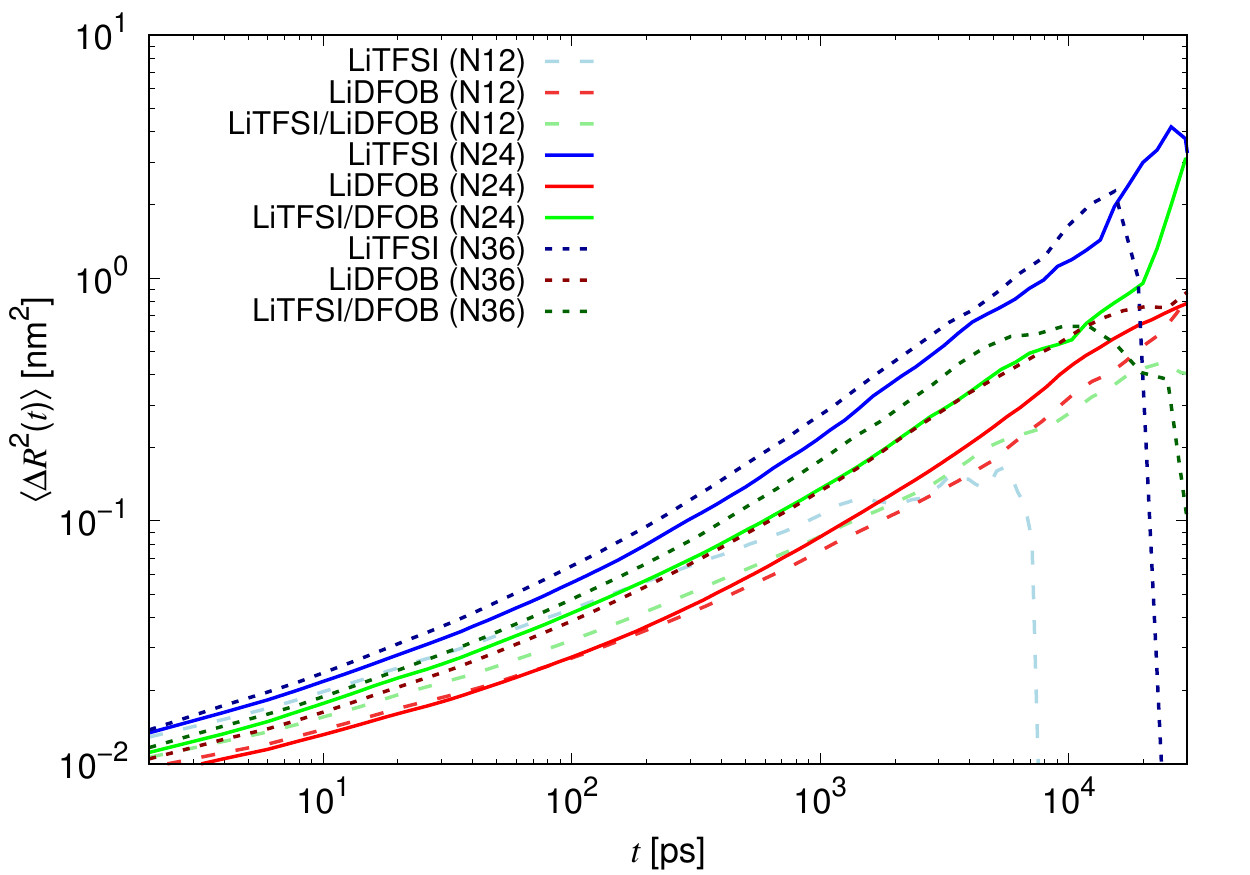}
		\caption{The MSDs of lithium ions exclusively bound to anions for time $t$ $\langle \Delta R^2_\mathrm{Li@anion}(t)\rangle$ are shown. The decrease of the MSDs at the end to zero indicate that there are no ions which are bound exclusively to anions for larger times $t$.}
		\label{SI:fig_msd_Li@anion} 
	\end{figure}
	
\clearpage
	
	\section{Network defects}

	\begin{figure}[h]
		\centering
		\includegraphics[width=\linewidth]{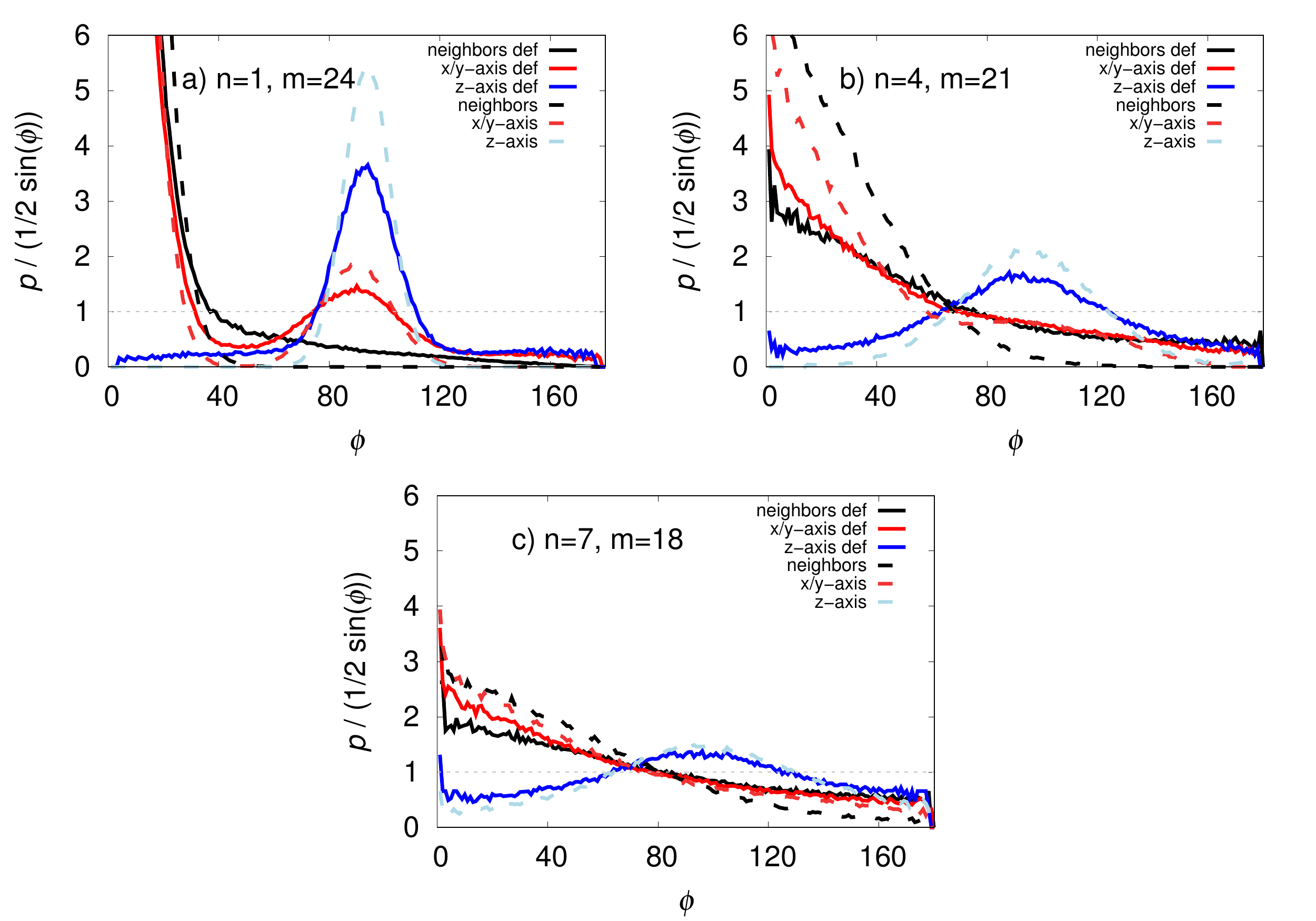}
		\caption{The probability distribution of the angles between PEO chains and unit vectors of the individual spatial dimensions and the angles between a vector in a PEO chain and a vector in a neigbouring PEO chain devided by $\frac{1}{2}\sin(\phi)$ are shown for a system with a PEO chain length of 24 monomers. Systems with network defects (solid line) and without network defects (dashed line) are compared. a) Shows the probability distribution calculated from the first EO in chain to the last EO in chain, b) shows the probability distribution calculated from the 4th EO in chain to the 21th EO in chain and c) shows the probability distribution calculated from the 7th EO in chain to the 18th EO in chain.}
		\label{SI:fig_angles_comparison}
	\end{figure}

	\begin{figure}[h]
		\centering
		\includegraphics[width=0.7\linewidth]{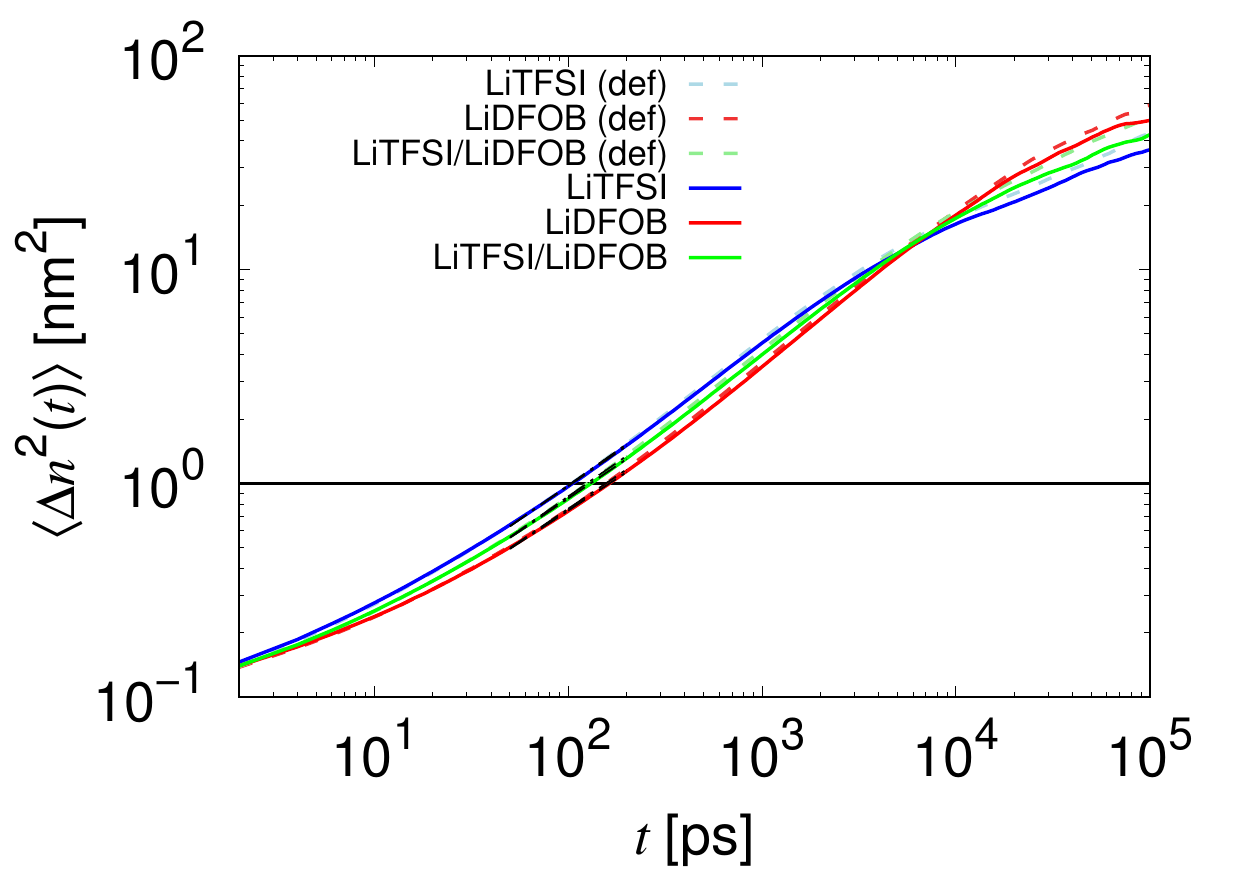}
		\caption{The one dimensional MSD $\langle \Delta n^2(t)\rangle$ of lithium ions along the polymer backbone is shown for systems with 24 monomers per chain with and without network defects. The black dashed lines are the used fits to calculate $\tau_\mathrm{1,eff}$.}
		\label{SI:fig_MSD_linear_def}
	\end{figure}

	\begin{figure}[h]
			\centering
		\includegraphics[width=\linewidth]{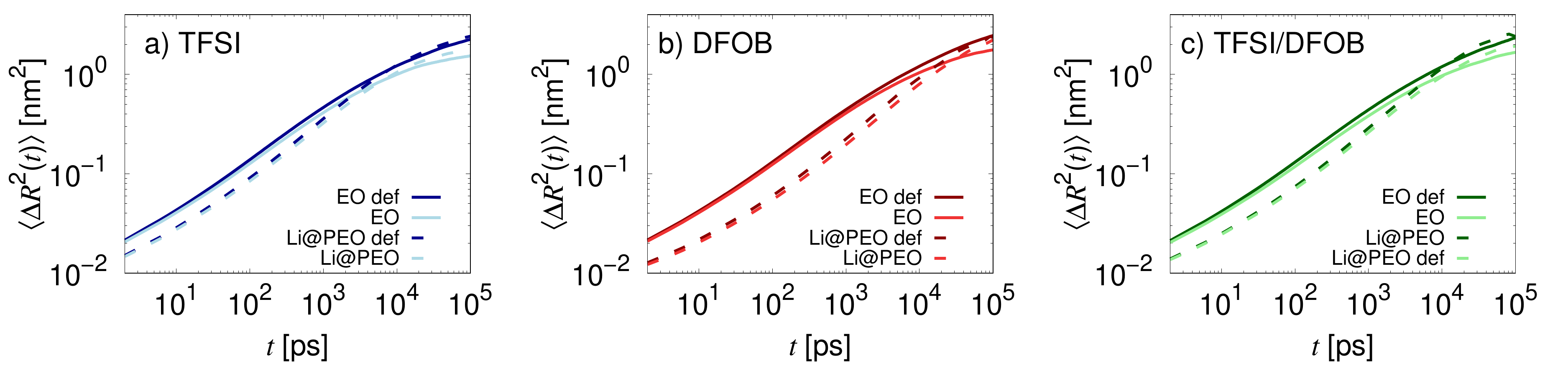}
		
		\caption{The mean squared displacement $\langle\Delta R_\mathrm{EO}^2(t)\rangle$ of the ether oxygens of PEO as a function of time $t$ and mean squared displacement $\langle\Delta R_\mathrm{Li@PEO}^2(t)\rangle$ of the lithium ions that are bound to a specific PEO chain at least for time $t$ as a function of time $t$ for a) TFSI, b) DFOB and c) TFSI/DFOB mixture are shown for a system with $N=24$ monomers per PEO chain. Systems with and without network defects are compared. Lithium ions are also considered as bound, if they jump to other anions or PEO chains and jump back to the specific chain again after a few picoseconds.}
		\label{SI:fig_msd_polymer_def}
	\end{figure}	
\clearpage

\printbibliography